\documentclass[usegraphicx,useAMS,usenatbib]{mn2e}

\usepackage{epsfig}
\usepackage{multirow}
\usepackage{amssymb}        

\newcommand{\gsim}{\gtrsim} 
\newcommand{\lsim}{\lesssim} 

\newcommand{\ergs}{{\rm erg\,s^{-1}}}

\newcommand{\Hz}{{\rm Hz}}

\newcommand{\mum}{\mu{\rm m}}

\newcommand{\hMpc}{\,h^{-1}{\rm Mpc}}

\def\g{{\scshape galform}}
\def\gr{{\scshape grasil}}

\begin{document}

\title[Clustering of sub-mm galaxies]
{Modelling the dusty universe II: The clustering of submillimetre-selected galaxies}

\author[Almeida et~al.]{
\parbox[t]{\textwidth}{
\vspace{-1.0cm}
C.\,Almeida$^{1}$,
C.\,M.\,Baugh$^{2}$,
C.\,G.\,Lacey$^{2}$
}
\\
$^{1}$Key Laboratory for Research in Galaxies and Cosmology,
Shanghai Astronomical Observatory, Chinese Academy of Sciences, \\
Nandan Road 80, Shanghai 200030, China.\\
$^{2}$Institute for Computational Cosmology, Department of Physics,
University of Durham, South Road, Durham, DH1 3LE, UK.\\
}

\maketitle
\begin{abstract}
We combine the \g~semi-analytical model of galaxy formation, which
predicts the star formation and merger histories of galaxies, the
\gr~spectro-photometric code, which calculates the spectral energy
distributions (SEDs) of galaxies self-consistently including
reprocessing of radiation by dust, and artificial neural networks
(ANN), to investigate the clustering properties of galaxies selected
by their emission at submillimetre wavelengths (SMGs). We use the
Millennium Simulation to predict the spatial and angular distribution
of SMGs. At redshift $z=2$, we find that these galaxies are strongly
clustered, with a comoving correlation length of $r_0 = 5.6 \pm
0.9\,h^{-1}$Mpc for galaxies with 850$\mum$ flux densities brighter
than 5 mJy, in agreement with observations. We predict that at higher 
redshifts these galaxies trace denser and increasingly rarer regions of 
the universe. We present the predicted dependence of the clustering on 
luminosity, submillimetre colour, halo and total stellar masses. 
Interestingly, we predict tight relations between correlation length 
and halo and stellar masses, independent of sub-mm luminosity.
\end{abstract}


\section{Introduction}
\label{section:intro}

The discovery of a population of high-redshift galaxies selected by their 
emission at submillimetre wavelengths (submillimetre galaxies; SMGs) has 
opened a new window on star formation in the high redshift Universe 
\citep[e.g.][]{smail97,hughes97,barger98,chapman00,blain02}. The
commonly held belief is that the submillimetre flux from these
galaxies is powered by prodigious star formation rates, which can reach
up $\sim 500 - 1000$ M$_{\odot}$ yr$^{-1}$ \citep{ivison00, smail02,
chapman05}.  The star formation is so intense that a substantial
fraction of the mass of the present day descendants of SMGs, bright
ellipticals, is thought to have been put in place during this phase
\citep{borys05,michalowski09}.  A further constraint on this picture
would come from an estimate of the mass of the dark matter haloes
which host SMGs. Measurements of the clustering of SMGs have so far
proved challenging \citep{scott02,blain04,borys05,scott06,weiss09}.
This situation has recently improved with the launch of the Herschel
telescope and will continue to get better with the commissioning of
the new SCUBA-2 camera on the James Clerk Maxwell Telescope.  In this
paper we present predictions for the clustering of SMGs using a galaxy
formation model set in the framework of the cold dark matter (CDM)
cosmology.

The self-consistent modelling of SMGs presents a number of challenges.
The sub-mm flux from a galaxy depends, often quite strongly, on a
number of galaxy properties and parameters of the dust model, such as
the star formation rate, the choice of the stellar initial mass
function (IMF), the dust extinction (which is driven by the optical
depth of the galaxy, which in turn depends on the metallicity of the
cold gas and the size of the galaxy), the nature and composition of
the dust grains and the thermal equilibrium temperature reached by the
dust grains when heated by starlight. Granato et~al. (2000) introduced
a hybrid model which combined a calculation of the star formation
histories of galaxies from the {\tt GALFORM} semi-analytical galaxy
formation model (Cole et~al. 2000) with the spectro-photometric code
{\tt GRASIL} (Silva et~al. 1998), which includes radiative transfer
through a two-phase dust medium and a self-consistent prediction of
dust temperatures. Using this model, a self-consistent calculation of
the dust emission from galaxies can be made \citep{baugh05,lacey08,
lacey10a}.

Constructing a galaxy formation model which can reproduce the observed
number counts of SMGs is relatively straightforward. It is more 
challenging to go a step further and to match the number counts and 
the redshift distribution of SMGs at the same time. 
The task becomes much more difficult if, at the same time as 
matching the properties of high redshift galaxies, the
model is also required to reproduce observations of the local galaxy
population. 
\cite{baugh05} argued that it is only possible to achieve
both of these goals by changing the slope of the IMF in episodes of
star formation triggered by galaxy mergers. By adopting a top heavy
IMF in starbursts, and by making the implied changes to the yield of
metals and the fraction of gas recycled from dying stars, Baugh
et~al. were able to reproduce basic properties of SMGs. At the same
time, the same model matches the observed luminosity function of
Lyman-break galaxies from $z=3$ to $z=10$ \citep[][]{lacey10b}, as well 
as being in good agreement with many observations of local galaxies. Nevertheless 
this model remains controversial and leads to conclusions which challenge 
the commonly accepted wisdom about SMGs. For example, Gonz\'alez et~al. (2010b)
investigated the nature of SMGs in the Baugh et~al. model and found
that the SMG phase is not responsible for the formation of a
significant amount of long lived stars. Here we present a further test
of the model by presenting predictions for the clustering of SMGs.

To date, there are suggestions that SMGs are strongly clustered, with
a correlation length that is substantially larger than that expected
for the dark matter at the typical redshift of SMGs (e.g. Blain
et~al. 2004).  However, the clustering measurements are currently
noisy as a result of the small volumes surveyed, with this scatter
being exacerbated by the strong clustering of the SMGs (Scott
et~al. 2002; 2006; Borys et~al. 2004; Weiss et al 2009). There is only
limited agreement between estimates of the angular clustering of SMGs, 
and poor agreement in turn between these measurements and the
clustering inferred in three dimensions. The first results from the
Herschel mission demonstrate the challenge of measuring the clustering
of SMGs. 
Cooray et~al. (2010) reported a detection of angular clustering in a
sample of galaxies selected at 250~$\mu$m from the Herschel HerMES
survey, which they estimate to have a similar redshift distribution to
the ``classical'' SMGs selected at 850~$\mu$m, while Maddox
et~al. (2010), on the other hand, found no evidence for angular
clustering for a galaxy sample selected in a similar way from the
Herschel ATLAS survey.  However, this situation is likely to improved
rapidly as the Herschel surveys increase in size and are analysed in
more detail.  Also, the SCUBA-2 camera is currently being installed
at the JCMT. The Cosmology Legacy survey using SCUBA-2 will produce a
map of $35$ deg$^{2}$ at $850\,\mu$m, substantially bigger than the
SHADES Half-Degree Extragalactic Survey.

There are currently few predictions for the clustering of SMGs. van
Kampen et~al. (2005) compiled predictions for the angular clustering
of SMGs from several groups. These calculations were phenomenological
and did not attempt to predict the sub-mm flux from galaxies. The
models were constrained by hand to match the SMG number counts.  Here
we make a direct prediction of which galaxies satisfy the selection
criteria to appear in an SMG sample. Gas dynamic simulations are
currently unable to provide meaningful predictions as the box sizes
used are too small to predict clustering robustly beyond a scale on
the order of a megaparsec. Furthermore, in many cases these
calculations stop at high redshift (again due to the small box size)
and so cannot be tested against the local galaxy population. By using
a semi-analytical approach, the computational resources can be devoted
to modelling the evolution of the dark matter component, allowing us
to use a representative cosmological volume.

In this paper we use the \g+\gr~model to populate the Millennium
Simulation of the evolution of structure in a cold dark matter
universe \citep{springel}. This simulation occupies a volume of
$500\,h^{-1}$Mpc on a side and contains more than 20 million dark
matter haloes at the present day. The {\tt CPU} required by \gr~makes
it impractical to compute the spectral energy distribution for every
galaxy directly in the Millennium Simulation. Instead, we apply a
novel technique based on artifical neural networks (ANN) which we
introduced in Paper I to populate the simulation with galaxies
\citep{ann}.

The paper is organised as follows. In Section 2 we give a brief 
summary of the \g+\gr~model and explain how the artificial neural 
network is implemented. We show how well the model can predict the
submillimetre luminosity of galaxies in Section 3. In Section 4 we 
present the predictions for the spatial and angular clustering of SMGs. 
The dependence of the clustering on selected galaxy properties is 
explored in Section 5. Finally, in Section 6, we present our conclusions.

\section{Model}
\label{section:model}

Here we give a brief summary of the semi-analytical 
galaxy formation model, \g (\S~\ref{section:galform}), 
the spectro-photometric model used to compute galaxy 
SEDs, \gr (\S~\ref{section:grasil}) and the artificial 
neural network (ANN) technique used to predict spectral 
properties for large samples of galaxies. Further details 
and tests of this approach can be found in \citet{ann}.

\subsection{The galaxy formation model: {\tt GALFORM}}
\label{section:galform}

In this paper we use the \g~galaxy formation model to follow the
fate of baryons in a $\Lambda$CDM universe. The general methodology
behind semi-analytical modelling is explained in the review by
\citet{baugh06}, and a more advanced overview of galaxy formation 
physics is given by \citet{benson10}. \g~was introduced 
by \citet{cole00}. Descriptions of subsequent extensions to the 
model are given in \citet{benson03, baugh05} and \citet{bower06}.

A summary of the model used in this paper, that of \citet{baugh05}, 
can be found in \citet{lacey08, lacey10a} and \citet{ann}, where 
the processes modelled are described and the parameters 
used to specify the model are listed.  
Two changes from the original Baugh et~al. set up are 
made as a result of the implementation of the model 
in the Millennium Simulation of Springel et~al. (2005). 
Firstly, the cosmological parameters of the Millennium 
are different from those adopted in the Baugh et~al. model.
\footnote{The Millennium Simulation adopts a flat $\Lambda$CDM
cosmology with a present-day matter density $\Omega_{\rm m} = 0.25$, a
cosmological constant of $\Omega_{\Lambda} = 0.75$, a Hubble constant
of $h \equiv H_0/(100$ km s$^{-1}$Mpc$^{-1}) = 0.73$ and a
perturbation amplitude given by the linear {\itshape rms} fluctuation
in spheres of radius $8\,h^{-1}$ Mpc of $\sigma_8 = 0.9$.  The
original Baugh et~al. model also assumes a flat $\Lambda$CDM cosmology
but with $\Omega_{\rm m} = 0.3$, $\Omega_{\Lambda} = 0.7$, $h = 0.7$
and $\sigma_{8}=0.93$.}  We have found that by adjusting the baryon
density parameter from the Baugh et~al. value of $\Omega_{\rm
b}=0.045$ to $\Omega_{\rm b}=0.033$ to give the same baryon fraction,
$\Omega_b/\Omega_{\rm m}$, as used in the original model, we obtain
similar predictions for the galaxy luminosity function to those
obtained in the original model.  Secondly, we use the merger histories
of the dark matter haloes extracted directly from the Millennium,
constructed using the prescription described in \citet{harker06}.

An important feature of the Baugh et~al. model, particularly for the
properties of galaxies selected by their dust emission, is the form of
the stellar initial mass function (IMF) adopted in different modes of
star formation. Bursts of star formation, which in this model are
triggered by certain types of galaxy merger, are assumed to produce
stars with a top-heavy IMF, where d$N/$d$ \ln m \propto m^{-x}$ and
$x=0$.  Bursts are initiated by all major mergers (i.e. those in which
the mass in cold gas and stars of the accreted galaxy account for 30\%
or more of the primary's mass) and by minor mergers in which the
accreted satellite makes up at least 5\% of the primary's mass and
where the primary is gas rich (defined as 75\% of the primary mass
being in the form of cold gas; these figures are model parameters).
Quiescent star formation in galactic disks is assumed to produce stars
according to a solar neighbourhood IMF, the \citet{kennicutt} IMF,
with $x=0.4$ for $m < M_{\odot}$ and $x=1.5$ for $m > M_{\odot}$.

The adoption of a top-heavy IMF in starbursts is the key to
reproducing the observed number counts and redshift distribution of
faint sub-mm galaxies (\citealt{baugh05,swinbank}).  While this choice
is controversial, a variety of observational evidence suggests that in
some environments the IMF may have a higher proportion of high-mass
stars than in the solar neighbourhood IMF \citep[see the review
by][]{elmegreen}. Moreover, the semi-analytical model is ideally
placed to investigate the consequences for other predicted properties
of assuming a top-heavy IMF in bursts.  A number of predictions have
been found to be in better agreement with observations following the
use of different IMFs in the burst and quiescent modes of star
formation, such as the metallicities of intra-cluster gas and of stars
in early-type galaxies (\citealt{nagashima05a,nagashima05b}). The
precise form of the IMF is not important so long as a higher
proportion of high mass stars are produced than would be the case with
a solar neighbourhood IMF. Similar predictions would be obtained for
an IMF with a standard slope which is truncated below a few solar
masses.  With a larger fraction of massive stars produced relative to
the Kennicutt IMF, more energy is radiated in the UV and larger
amounts of dust are produced due to the enhanced yield of metals.

In the next subsection we describe the \gr~spectro-photometric 
code which generates a spectral energy distribution for each galaxy 
across a wide range of wavelengths. \g~itself makes an independent
calculation of the spectral energy distribution (SED) of starlight, including  
a model for dust extinction which is described in \citet{cole00}. 
This calculation gives similar results to those obtained with \gr\ 
at optical wavelengths. The \g~calculation of the V-band luminosity 
weighted age and optical depth are used as inputs to the ANN.

\subsection{The spectro-photometric model: {\tt GRASIL}}
\label{section:grasil}

To accurately predict the SEDs of galaxies, from the far-UV to the
radio, we use the spectro-photometric code \gr~\citep{silva98}.  This
code computes the stellar emission, absorption and emission of
radiation by dust, and radio emission powered by massive stars
\citep{bressan02}. \gr~carries out an accurate treatment of the
extinction and reprocessing of starlight by dust.

The combination of \g~and \gr~was described by \cite{granato00} and
has been exploited in a series of papers
\citep{baugh05,lacey08,lacey10a}.  The semi-analytical model calculates
the star formation and metal enrichment history for each galaxy,
including the contribution from starbursts.  \g~also predicts the
scalelengths of the disk and bulge components of each galaxy, as
described in Cole et~al. (2000) and tested by \cite{almeida07} and
\cite{gonzalez09}, and the cold gas mass (as compared against
observations by \cite{power10} and \cite{kim10}).  The dust is
modelled as a two-phase medium, with a diffuse component and dense
molecular clouds. The mass split between these components is a model
parameter. In the Baugh et~al. model, 25 per cent of the dust is
assumed to be in the form of dense clouds. Stars form within molecular
clouds and escape on a timescale $t_{\rm esc}$, which is another model
parameter; in \citet{baugh05}, a value of $t_{\rm esc} = 1$ Myr is
adopted in both quiescent and burst modes of star formation.  The
extinction of starlight by dust clouds depends on the star's age
relative to the escape time.  High mass stars, which typically
dominate the emission in the UV, spend a significant fraction of their
comparatively short lifetimes within molecular clouds.  \gr~calculates
the radiative transfer of starlight through the dust and
self-consistently solves for the temperature distribution of the dust
grains at each point in the galaxy, based on the local radiation
field.  The temperature distribution of the grains is then used to
calculate the dust emission. The composition and size of the dust
grains are chosen to match the properties of the local ISM: a mixture
of graphite and silicate grains, as well as polycyclic aromatic
hydrocarbon (PAH) molecules.  The effects of temperature fluctuations
in very small grains and PAH molecules are taken into account. Emission
from PAHs is calculated using the cross-sections of
\citet{li01}. Radio emission from ionised HII regions and synchrotron
radiation is included as in \citet{bressan02}.

The \gr~model has been calibrated against local observational data for
normal and starburst galaxies \citep{bressan02, vega05, panuzzo07,
schurer09}.  A limitation of \gr~is that it assumes axisymmetric
distributions for the gas and dust in starburst galaxies. There is
observational evidence of more complex geometries and extraplanar dust
in some galaxies \citep[see for example][]{engelbracht06}.  This could
be problematic if this dust absorbed and emitted a significant fraction
of radiation. However, there is little observational evidence for
this. Furthermore, observations of nearby starbursts reveal that most
of the absorption and emission of radiation by dust takes place in a
compact region of size $\sim 1$ kpc or less.  \gr~has been shown to 
accurately predict the SEDs of both quiescent and starburst 
galaxies \citep{silva98, bressan02}.

\subsection{The artificial neural network approach to predicting 
galaxy luminosities}
\label{section:ann}

The \gr~code provides an accurate calculation of the absorption and
reemission of radiation by dust, predicting the SED of a galaxy from
the far-UV to radio. However, \gr\ is extremely CPU intensive,
requiring several minutes to compute the SED for a single galaxy,
which prohibits its direct application to extremely large numbers of
galaxies.  In \citet{ann} we introduced a new technique based on
artificial neural networks (ANN) which can be used to rapidly predict
SEDs using a small set of galaxy properties as input, once the ANN has
been trained on a relatively small number ($\sim 2000$) of galaxies
with SEDs computed using \gr. We demonstrated that, in the majority of
cases, this method can predict the luminosities of galaxies to within
10 per cent of the values computed directly using \gr. We employ the
same approach in this paper.  The general methodology behind the ANN
is set out in detail in \citet{ann}, so we give only a brief summary
here. \cite{silva10} recently published a complementary approach in
which the explicit calculation of emission by dust within \gr~is
replaced by an ANN.

Artificial neural networks are mathematical models designed to
replicate the behaviour of the human brain. They are similar to their
biological counterparts in the sense that ANNs consist of simple
computational units, neurons, which are interconnected in a
network. The neurons are usually organized in layers: an input layer,
one or more hidden layers and an output layer. Each neuron has a
weight associated with it. In this paper we will use a multilayer,
feed-forward network. The ANN is trained using a sample of galaxies
for which \gr~has been run to compute spectra. During the training
process, the neural network is presented with a set of inputs,
comprised of selected galaxy properties, and associated outputs, in
our case the luminosity at different wavelengths. The network weights
are adjusted in order to reproduce, as closely as possible, the
desired output from the given set of inputs. In summary: (i) we start
with an untrained net (random weights); (ii) determine the output for
a given input; (iii) compute the discrepancy or error between the
predicted and the target output; and (iv) adjust the weights in order
to reduce this error. To adjust the weights we use the resilient
backpropagation learning algorithm \citep{rprop}.

As in \citet{ann}, we use 12 galaxy properties predicted by \g~as
input to the ANN: the total stellar mass, the stellar metallicity, the
unextincted stellar bolometric luminosity, the disk and bulge
half-mass radii and the circular velocities measured at these two
radii, the V-band weighted age, the optical depth of dust extinction
in the V-band, the metallicity of the cold gas, the mass of stars
formed in the last burst and the time since the start of the last
burst of star formation.  \citet{ann} showed that the performance of
the ANN is greatly improved if we predict only one output property,
the luminosity in a single band-pass, instead of predicting the full
SED of the galaxy (which typically covers 456 wavelength bins for a
standard \gr\ SED).  We then train the ANN separately for each band
required.  Here, we follow the same approach: we train two separate
networks, one for each of the sub-mm wavelengths at which we want to
predict luminosities ($450\,\mu$m and $850\,\mu$m).  The network
configuration adopted has 12 neurons in the input layer, two hidden
layers with 30 neurons each and one output neuron.  We also found, in
order to maximise the accuracy of the ANN predictions, that it was
necessary to train the ANN at each redshift of interest, and to train
separately for galaxies whose star formation is dominated by
starbursts or quiescent star formation in disks.

\subsection{The performance of the ANN at sub-mm wavelengths}
\label{section:performance}

\begin{figure}
{\epsfxsize=8.truecm
\epsfbox[18 144 592 718]{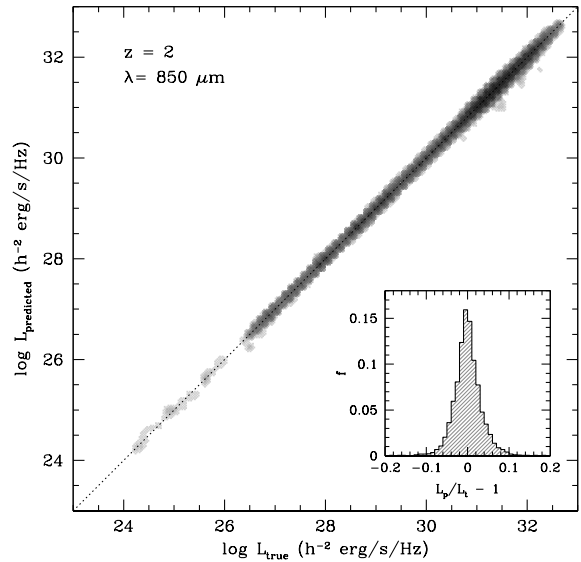}}
\caption
{
Comparison between the ANN predicted luminosity, $L_{\rm predicted}$,
and the true luminosity calculated using \gr, $L_{\rm true}$, 
at $850\,\mu$m in the observer frame for galaxies at $z=2$.  
Note that a flux of 1~mJy at $z=2$ corresponds to a luminosity 
$L_{\nu} = 4.75\times 10^{31} h^{-2} \ergs\Hz^{-1}$. The shading 
shows the distribution of galaxies in the training sample.  
In the inset, we plot the error distribution of the predicted 
luminosities, as given by $L_{\rm predicted}/L_{\rm true} - 1$, 
normalized to unit area.
}
\label{fig:performance.lplo}
\end{figure}

\begin{table*}
\begin{center}
  \begin{tabular}{cccccc}
  \hline
 & & \multicolumn{2}{c}{$450\,\mu$m} & \multicolumn{2}{c}{$850\,\mu$m} \\
 Redshift & Sample & $\varepsilon_L$ & $P_{|e|<10 \,{\rm per\,cent}}$  & $\varepsilon_L$ & $P_{|e|<10\,{\rm per\,cent}}$  \\
 \hline
 \multirow{2}{*}{z = 0.1}
 & Quiescent	& 0.04 & 97.1 & 0.04 & 98.6 \\
 & Burst 	& 0.07 & 88.3 & 0.13 & 87.0 \\
\multirow{2}{*}{z = 0.5}
 & Quiescent	& 0.04 & 96.9 & 0.04 & 98.5\\
 & Burst 	& 0.05 & 94.8 & 0.05 & 96.7 \\
\multirow{2}{*}{z = 1}
 & Quiescent	& 0.06 & 94.9 & 0.04 & 97.0 \\
 & Burst 	& 0.08 & 90.1 & 0.07 & 93.2 \\
\multirow{2}{*}{z = 2}
 & Quiescent	& 0.04 & 97.2 & 0.04 & 97.5 \\
 & Burst 	& 0.05 & 92.1 & 0.05 & 95.1\\
\multirow{2}{*}{z = 3}
 & Quiescent	& 0.05 & 96.0 & 0.03 & 97.8 \\
 & Burst 	& 0.05 & 93.4 & 0.03 & 98.2 \\
\multirow{2}{*}{z = 4}
 & Quiescent	& 0.04 & 97.0 & 0.03 & 98.5 \\
 & Burst 	& 0.05 & 93.2 & 0.03 & 97.9 \\
  \hline
  \end{tabular}
   \caption{ Statistics of the error distribution associated with the
ANN prediction of $450\,\mu$m and $850\,\mu$m observer-frame
luminosities at selected redshifts. The statistics are computed using
only galaxies with sub-mm fluxes brighter than 1~mJy in that band.
Column 1 gives the redshift, column 2 specifies whether the galaxy
sample is made up of galaxies forming stars quiescently or starbursts;
columns 3 and 4 give $\varepsilon_L$ (the root mean squared error
defined by Eq.~\ref{eq:varepsilon}) and $P_{|e|<10\,{\rm per\,cent}}$
(percentage of galaxies with predicted luminosities within 10\% of the
true value), respectively, for the $450\,\mu$m predictions. For
$850\,\mu$m selected galaxies, the same information is shown in
columns 5 and 6.  }
  \label{tab:performance}
  \end{center}
 \end{table*}

We now demonstrate the how well the ANN performs when 
predicting galaxy luminosities in the sub-mm. Test at other wavelengths 
were presented in \cite{ann}.  

In Fig.~\ref{fig:performance.lplo} we plot the comparison between the
observer frame luminosity in the SCUBA $850\,\mu$m band predicted by
the ANN for $z=2$ galaxies and the true values calculated directly
using \gr.  In this plot we include all galaxies regardless of their
classification as quiescent or starburst galaxies.
Fig.~\ref{fig:performance.lplo} shows that there is excellent
agreement between the luminosities predicted by the ANN and the true
values, with most of the predicted luminosities being within 10 per
cent of the \gr~result (inset).  Some statistics quantifying the error
distribution at different redshifts are summarized in
Table~\ref{tab:performance}, for galaxies brighter than 1~mJy in the
corresponding band (either 450 or 850~$\mum$).  Here, the root mean
squared logarithmic error, $\varepsilon_L$, is defined by:
\begin{equation}
\label{eq:varepsilon}
 \varepsilon_L = \sqrt{1/n \sum^n [\ln (L_{\rm predicted}/
L_{\rm true})]^2} \, ,
\end{equation}
where $n$ is the number of galaxies considered. The quantity 
$P_{|e|<10 {\rm per\,cent}}$ is defined as the percentage of 
galaxies with predicted luminosities which lie within 10 per 
cent of the true values.  For quiescent SMGs, we are able 
to reproduce the luminosities of more than 95 per cent of galaxies 
with an accuracy of 10 per cent or better, for the redshift range 
considered, at both $450\,\mu$m and $850\,\mu$m. As shown by 
\citet{ann}, the performance of the ANN for burst galaxies is
somewhat poorer, which is a consequence of the wide range of spectra 
seen in bursts and the difficulty the ANN experiences in reproducing 
this variety. Nonetheless the technique returns more than 90\% of
predicted sub-mm luminosities within 10 per cent of the true values.  
It should be noted that at $z=4$, the highest redshift considered, 
the observer-frame $850\,\mu$m luminosity probes the rest-frame 
$170\,\mu$m, which is approaching the peak in the dust emission 
spectrum (typically around $100\,\mu$m). 

One important feature of the error distribution is shown in the inset
of Fig.~\ref{fig:performance.lplo}. The distribution of luminosity
errors predicted by the ANN appears to be Gaussian. Furthermore, we
find that there is no correlation of the error with luminosity or
other galaxy properties.  This suggests that any sample of SMGs built
using the ANN method will have errors which are decoupled from the
structural and photometric properties of the galaxy sample.

\section{Clustering of sub-mm galaxies}
\label{section:clustering}

The clustering of galaxies is an important constraint on the masses of
their host dark matter halos, and hence on theoretical models of
galaxy formation, as it depends upon how various physical processes
vary with halo mass. In this section we present the model predictions
for the clustering of galaxies selected by their flux at sub-mm
wavelengths. In \S3.1, we contrast the spatial distribution of
galaxies with that of dark matter haloes. In \S3.2, we define the
two-point spatial correlation function. We demonstrate that our
clustering predictions are insensitive to errors in the galaxy
luminosities predicted by the ANN in \S 3.3. We present the
predictions for clustering in real space and redshift space in \S 3.4
and \S 3.5 respectively. The evolution of the correlation function is
presented in \S 3.6 and the angular correlation function is shown in
\S 3.7.

\subsection{The spatial distribution of SMGs} 

\begin{figure*}
{\epsfxsize=16.truecm
\epsfbox[0 10 579 560]{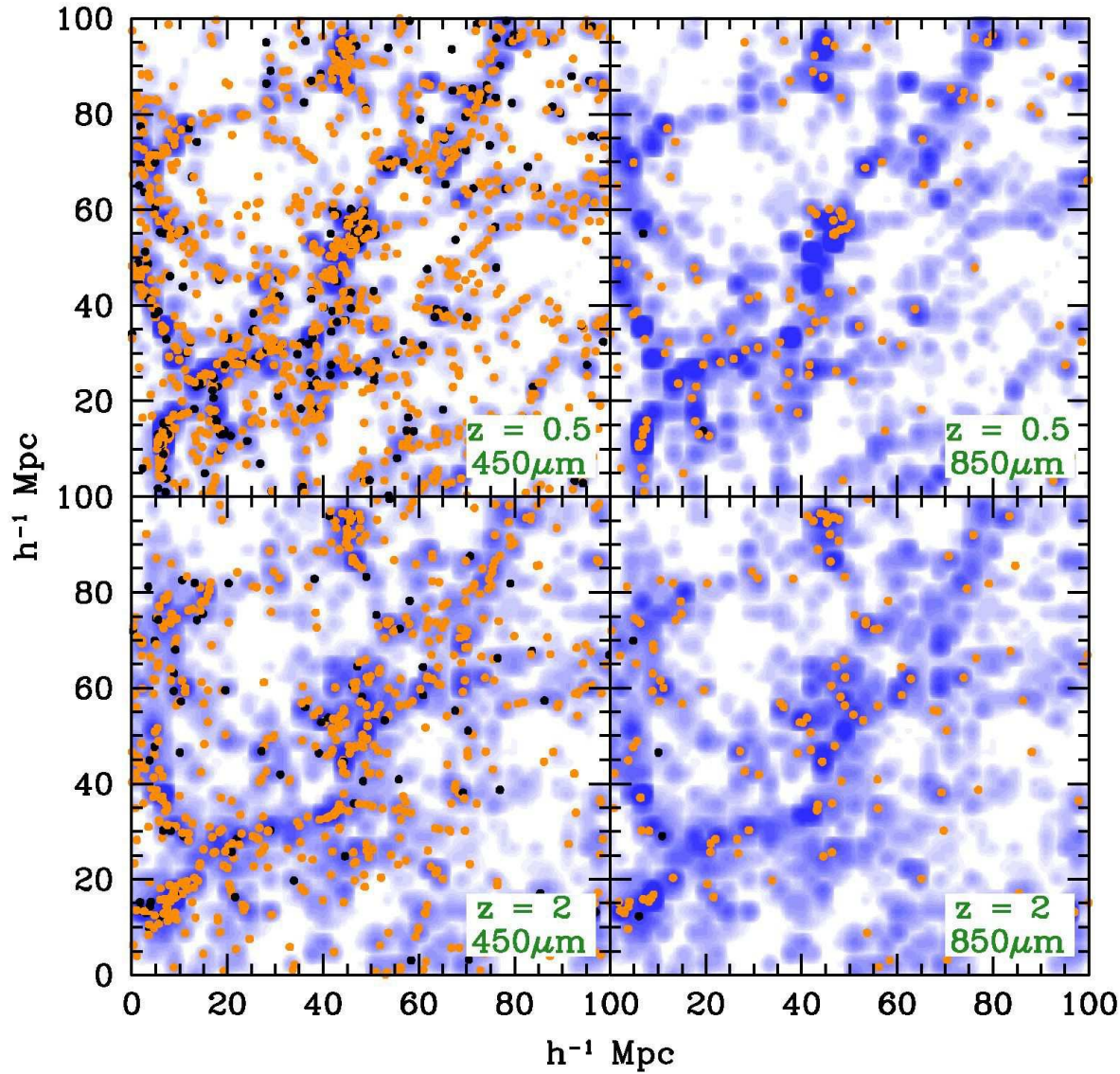}}
\caption{Image of the simulated spatial distribution of SMGs and dark
  matter haloes at $z=0.5$ (top panels) and $z=2$ (bottom panels). The
  panels display a slice $100\,h^{-1}$ Mpc wide with a depth of
  $20\,h^{-1}$ Mpc (in comoving coordinates). The width of this slice 
  corresponds to angular scales of $4.3^\circ$ and $1.6^\circ$ at $z=0.5$ and $z=2$
  respectively. The dark matter haloes are plotted in blue, with the
  darker shading corresponding to regions of higher projected halo mass 
  density.
  Galaxies with $S_{450\mu m}$ and $S_{850\mu m} \ge 1$ mJy are
  represented by the orange dots. Brighter galaxies with 
  $S \ge 5$ mJy are shown by the black dots. 
  The left hand panels show galaxies selected by their $450\,\mu$m flux, 
  while the right hand side show galaxies selected by their emission at 
  $850\,\mu$m.}
\label{fig:spatial}
\end{figure*}

Before discussing the predictions for the two-point correlation
function of SMGs, we first gain a visual impression of their spatial
distribution. Fig.~\ref{fig:spatial} shows dark matter haloes and
sub-mm galaxies in a slice taken from the Millennium Simulation. The
slice measures $100\,h^{-1}$ Mpc across and $20\,h^{-1}$ Mpc thick in
comoving units. The upper panels show haloes and galaxies at $z=0.5$
and the lower panels show them at $z=2$.  Dark matter haloes are shown
by the blue shading. The intensity of the shading is proportional to
the total halo mass within each pixel.  We show sub-mm galaxies
selected at $450\,\mu$m and $850\,\mu$m with fluxes brighter than 1
mJy and 5 mJy.  At a given flux limit, the $450\,\mu$m sources are
more numerous than the $850\,\mu$m sources.  The $850\,\mu$m sources
brighter than $1$mJy tend to trace out the more massive dark matter
haloes and hence are expected to be biased tracers of the dark matter
distribution.  The full Millennium box ($500 h^{-1}$Mpc across)
subtends an angle of 7.5 degrees at $z=2$.  To put this into context,
we note that the SCUBA-2 Cosmology Legacy Survey
(SCLS)\footnote{http://www.jach.hawaii.edu/JCMT/surveys/Cosmology.html}
aims to map around 35~$\deg^2$ at 850~$\mum$ in patches up to 3~$\deg$
across, and 1.3~$\deg^2$ at 450~$\mum$ in regions up to 0.5~$\deg$
across. The nominal $5\sigma$ flux limits will be 3.5~mJy at
850~$\mu$m and 2.5~mJy at 450~$\mu$m.  However, source confusion may
result in the flux limits for reliable source identification being
somewhat brighter than this; using the standard 20 beams per source
criterion for confusion \citep[e.g.][]{lacey08} together with the
counts predicted by the model, we expect confusion to become important
around 3.0~mJy at 850~$\mu$m and 5.3~mJy at 450~$\mu$m.

\subsection{The two-point correlation function}
\label{section:two_point}

To quantify the clustering of the galaxy distribution we use 
the two-point correlation function, $\xi(r)$, which gives the 
excess probability, compared with a random distribution, of 
finding two galaxies at a separation $r$: 
\begin{equation}
\label{eq:prob1}
 \delta P(r) = \bar{n}^2\,[1 + \xi(r)]\,\delta V_1\,\delta V_2,
\end{equation}
where $\bar{n}$ is the mean space density of galaxies and the $\delta
V_{\rm i}$ are elements of volume. 
If $\xi(r) > 0$, then galaxies are more clustered than a random distribution. 
On the contrary, if galaxies have a tendency to avoid one other, 
then $\xi(r) < 0$.

The two-point correlation function of galaxies is shaped by two main
factors, which play different roles on different scales. On large
scales, the form of the correlation function is controlled by the
clustering of galaxies in distinct dark matter haloes (referred to as
the {\em two-halo term}), and the galaxy and dark matter correlation
functions have similar shapes but differ in amplitude (for an
illustration of this see Angulo et~al. 2008).  On smaller scales, up
to the size of the typical haloes which host galaxies, the form of the
correlation function is driven by the number and radial distribution
of galaxies within the same dark matter halo, (called the {\em
one-halo term}) \citep{benson00, seljak00}.

We calculate the two-point correlation function of sub-mm selected galaxies, 
using both real and redshift space coordinates.
We measure the correlation function using the standard estimator (e.g. 
Peebles 1980):
\begin{equation}
 \xi(r) = \frac{\langle DD(r) \rangle}{\frac{1}{2}\,N_{\rm gal}\,\bar{n}\,\Delta V(r)} - 1,
\end{equation}
where $\langle DD(r) \rangle$ is the number of distinct galaxy pairs with
separations between $r$ and $r + \Delta r$, $N_{\rm gal}$ is the total 
number of galaxies, and $\bar{n}$ is the mean number density of 
galaxies. $\Delta V(r)$ is the volume of a spherical shell of radius $r$ 
and thickness $\Delta r$. We are able to compute the volume of this shell 
analytically since we are dealing with galaxy pairs within a periodic 
simulation volume. The clustering signal is generated in redshift space 
using the distant observer approximation, by electing one axis to be 
the line of sight direction, and adding the suitably scaled peculiar 
velocity of the galaxy along this axis to its comoving position.

\subsection{The impact of luminosity errors on the predicted correlation function}

\begin{figure}
{\epsfxsize=8.truecm
\epsfbox[18 144 592 718]{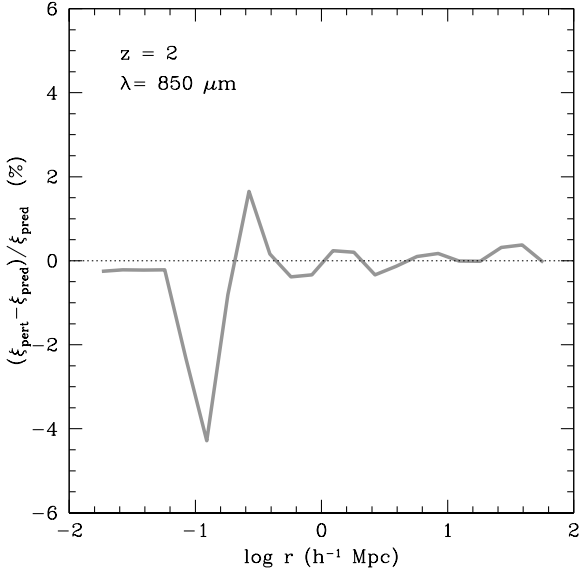}}
\caption{
  The impact of errors in the ANN predicted $850\mu$m luminosities 
  on the form of the real-space two-point correlation function. 
  The correlation function is measured for galaxies selected with fluxes 
  brighter than 1 mJy at $z=2$ ($\xi_{\rm pred}$). The galaxy fluxes are 
  then perturbed using the error distribution of the ANN 
  $850\,\mu$m luminosities and the correlation function is re-measured 
  for the new sample of galaxies brighter than 1 mJy ($\xi_{\rm pert})$. 
  The plot shows the maximum deviation of $\xi_{\rm pert}$ from 
  $\xi_{\rm pred}$ on constructing 20 different perturbed samples, 
  expressed as a percentage. The spike around $\log(r/h^{-1} {\rm Mpc}) 
  \approx -0.8$ is caused by noise due to the small number of galaxy
  pairs at that separation.}
\label{fig:random}
\end{figure}

We now look at the impact of errors on the ANN-predicted luminosities
on the amplitude and shape of the two-point correlation function. 
First, the correlation function, $\xi_{\rm pred}(r)$, is 
measured using the ANN-predicted observer-frame 850~$\mu$m luminosities  
for a sample brighter than some flux limit, in this case $1$~mJy. 
Next, these luminosities are perturbed by drawing an offset from 
the distribution of errors expected for this band 
(see inset of Fig.~\ref{fig:performance.lplo} 
and Table~\ref{tab:performance}). 
A new flux limited sample is constructed, 
which will contain some galaxies which were not included in the 
initial, unperturbed sample, because their fluxes have been boosted. 
Moreover some galaxies which 
made it into the original sample will no longer be included after their 
luminosities have been perturbed. 
We then repeat the measurement of the correlation function for 
this new sample, resulting in the estimate $\xi_{\rm pert}(r)$. 
Comparing $\xi_{\rm pert}(r)$ with
$\xi_{\rm pred}(r)$ provides an estimate of how $\xi(r)$ is affected 
by the errors in the ANN-predicted luminosities.  In
Fig.~\ref{fig:random} we plot the maximum deviation of the ratio
$\xi_{\rm pert}(r)/\xi_{\rm pred}(r)$ using 20 different $\xi_{\rm
pert}(r)$ measurements for galaxies at $z=2$ (i.e. after perturbing 
the galaxy luminosities).  
This plot shows that $\xi_{\rm pert}(r)/\xi_{\rm pred}(r)$ differs 
from unity by at most 4\%, indicating that the clustering predictions 
are essentially unaffected by the errors in the ANN luminosities. 
At $z=0.1$, where the performance of the ANN is poorer for
galaxies undergoing a burst of star formation than it is at $z=2$, 
the ratio $\xi_{\rm pert}(r)/\xi_{\rm pred}(r)$ still deviates 
from unity by less than 10\%.

\subsection{The real space correlation function}
\label{sec:xi-real-space}

\begin{figure}
{\epsfxsize=8.truecm
\epsfbox[18 144 592 718]{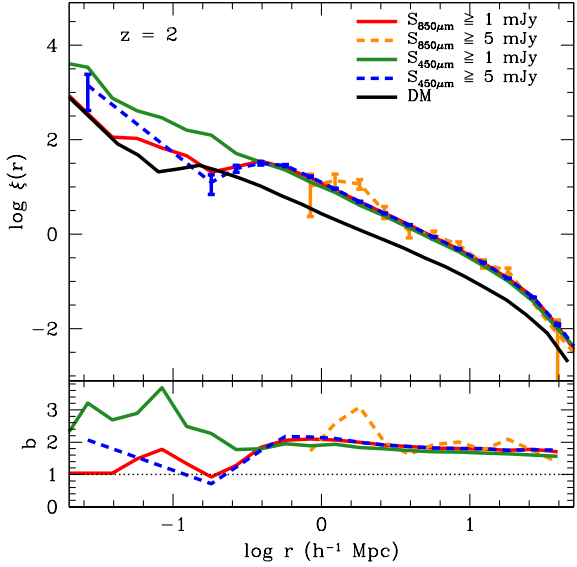}}
\caption{
  The real space two-point correlation function of sub-mm
  selected galaxies in comoving coordinates at $z=2$. The blue and orange
  dashed lines show the predicted correlation functions for galaxies
  with fluxes brighter than 5~mJy at 450 and 850~$\mu$m respectively,
  while the green and red lines show the corresponding results for a
  flux limit of 1~mJy. The black line shows the correlation function
  of dark matter. The errorbars show the $1\sigma$ Poisson errors
  derived from the number of pairs in each bin of radial separation, and
  are only shown for those samples with the lowest total number of
  galaxies. In the bottom panel, we plot the galaxy bias, $b(r)$, as a
  function of scale for the four samples of sub-mm selected galaxies.
  This is obtained by taking the square root of the ratio of the galaxy 
  correlation function to the measured dark matter correlation function. 
  }
\label{fig:clustering.real.z2}
\end{figure}

In real space, the cartesian coordinates of the SMGs within the
simulation box are used to calculate pair separations. 
Fig.~\ref{fig:clustering.real.z2} (top panel) shows the real space correlation
function for sub-mm selected galaxies, for both $\lambda = 850\,\mu$m
and $\lambda = 450\,\mu$m, at redshift $z=2$. The transition between
the one-halo term and two-halo term occurs around $r\sim 0.6 h^{-1}$Mpc.  
In this plot we consider samples of galaxies selected to 
be brighter than 1 mJy or 5 mJy at both wavelengths. 
The black line shows the correlation function of the 
dark matter, $\xi_{\rm DM}$, which was measured using a  
randomly chosen subset of $10^6$ dark matter particles 
out of the ten billion particles in the Millennium Simulation.

As can be seen in Fig.~\ref{fig:clustering.real.z2}, galaxies do not
trace the mass distribution in the Universe, because the efficiency 
of galaxy formation depends on halo mass \citep[e.g.][]{eke04}.
The difference between the clustering of galaxies and the underlying 
dark matter is quantified in terms of the galaxy clustering bias, $b$:
\begin{equation}
\label{eq:bias}
 b(r) = \left( \frac{\xi_{\rm gal} (r)}{\xi_{\rm DM}(r)}
 \right)^{1/2}.
\end{equation}
Numerical simulations have demonstrated that the galaxy bias is a
function of scale (Smith, Scoccimarro \& Sheth 2007; Angulo et al
2008).  This scale dependence weakens on large scales, and the galaxy
bias is typically approximated as a constant. We plot the bias of SMGs
in the lower panel of Fig.~\ref{fig:clustering.real.z2}. The plot
shows that at $z=2$ the bias factor is generally greater than unity,
approaching a roughly constant value of $b \approx 1.8$ for $r \gsim
2\hMpc$ for all of the samples shown.  In the case of galaxies
selected at $450\,\mu$m, there is a small but clear difference in the
bias predicted for bright and faint samples, with the bright galaxies
being the more strongly clustered.  At $850\,\mu$m the distinction is
less clear, due in part to the relatively low number density of
galaxies in the bright sample, which results in a noiser prediction.

The effective bias parameter on large scales can also be  
estimated analytically \citep{mo96, smt01}, using the mass 
function of haloes which host sub-mm galaxies, $N(z, M)$, 
(i.e. the product of the space density or mass function of 
dark matter haloes and the halo occupation distribution of 
SMGs) and the bias factor as a function of the halo mass 
$b(z, M)$ \citep[e.g.][]{baugh99}:
\begin{equation}
 b_{\rm eff}(z) = \frac{\int_M N(z, M')\,b(z, M')\,{\rm d}\ln M'}
{\int_M N(z, M')\,{\rm d}\ln M'}
\end{equation}
The integrals are taken over the full range of halo masses, with
$N(z,M) = 0$ for haloes which do not host SMGs.  To compute $b(z, M)$
we use the prescription outlined by \citet{smt01}. For galaxies at
$z=2$ with $S_{850\mu \rm m}\ge 5$~mJy, we find an effective bias of
$b_{\rm eff} = 2.3$, and $b_{\rm eff} = 2.1$ for sub-mm galaxies with
$S_{450\mu m}\ge 5$~mJy. These values are slightly larger than those
estimated from the simulation using Eq.~\ref{eq:bias}.  In fact,
similar differences between the analytical approach and simulations
have been observed in other studies \citep[see for example][]{gao05,
angulo09}.

On small scales, Fig.~\ref{fig:clustering.real.z2} shows that 
the effective bias takes on a range of values. 
The clustering on these scales is driven by the typical number 
of galaxy pairs within a common halo. The faint sample of galaxies 
selected at $450 \mu$m displays the strongest clustering on 
small scales. This sample contains the largest number of pairs 
within common haloes. In the case of the bright sample at $850 \mu$m, 
the low number density of galaxies makes it difficult to measure 
the correlation function on small scales. 

A convenient measure of the strength of clustering for different
galaxy samples is provided by the correlation length $r_0$, which we
can define in a robust way as the pair separation at which the 
correlation function becomes unity: 
\begin{equation}
 \xi(r_0) = 1.
\label{eq:r0}
\end{equation}
Applying this definition to model galaxies selected at 850~$\mu$m, we
find $r_0 = 5.6 \pm 0.9 \hMpc$ for $S_{850\mu \rm m}\ge 5$~mJy, and $r_0 =
5.38 \pm 0.02 \hMpc$ for a fainter sample with $S_{850\mu \rm m}\ge
1$~mJy. We also find that $450\,\mu$m selected galaxies are less
clustered than $850\,\mu$m selected galaxies at the same flux 
limit: we obtain $r_0 = 5.38 \pm 0.02 \hMpc$ and $4.99 \pm 0.01 \hMpc$
respectively for $S_{450\mu m}\ge 5$~mJy and 1~mJy. SMGs with
$S_{450\mu m}\ge 5$~mJy display a similar two-point correlation
function to that of S$_{850\mu m}\ge 1$ mJy (as we will see later,
this is mainly a consequence of the fact that the median S$_{450\mu
m}$/$S_{850\mu m}$ colour is approximately 3).

An alternative way to define the clustering length $r_0$, commonly
used in observational studies, is to fit the correlation function with
a power law:
\begin{equation}
 \xi(r) = \biggl(\frac{r}{r_0}\biggr)^\gamma.
\label{eq:xi}
\end{equation}
For optical galaxies at $z\sim 0$, this is found observationally to
provide a good fit for $0.1 < r < 10\,h^{-1}$Mpc with $\gamma$ close
to -1.8 \citep[e.g.][]{norberg01,zehavi05}. The two definitions of $r_0$
(Eqs.~\ref{eq:r0} and~\ref{eq:xi}) are obviously equivalent only
if $\xi(r)$ really is a power law. If $\xi(r)$ actually has a 
more complicated dependence on $r$, then the value of $r_0$ obtained 
by fitting a power law will depend on the range of $r$ over which 
the fit is performed, and on the errors on the measurements at different $r$.
If we fit $\xi(r)$ for our model SMGs at $z=2$ with a power law over
the range $0.1 < r < 10\,h^{-1}$Mpc, we find $\gamma = -1.8 \pm 0.2$
and a correlation length of $r_0 = 6 \pm 1 \,h^{-1}$Mpc for
$S_{850\mu m}\ge 5$~mJy, and $\gamma = -1.61 \pm 0.01$ and $r_0 = 5.21
\pm 0.05 \,h^{-1}$Mpc for $S_{850\mu m}\ge 1$~mJy.  For galaxies with
$S_{450\mu m}\ge 5$~mJy, we find a slope similar to that of
$850\,\mu$m galaxies, $\gamma = -1.62 \pm 0.01$, and $r_0 = 5.20 \pm
0.07\,h^{-1}$Mpc. The values of $r_0$ obtained from the power law fit
are thus close to the values obtained from the more general definition 
(Eq.~\ref{eq:r0}) in this case. However, we will use Eq.~\ref{eq:r0} 
to define $r_0$ unless stated otherwise.

Our predictions for the clustering of the brighter SMGs at 850~$\mum$
are in reasonable agreement with the observational estimate by
\citet{blain04} who, using a sample of 73 SMGs with $S_{850\mu m}\ge
5$~mJy and spectroscopic redshifts at $z \approx2-3$, inferred a
correlation length of $6.9 \pm 2.1\,h^{-1}$Mpc using a pair-counting
approach rather than a direct measurement of $\xi(r)$ (note this
measurement is discussed further in \S 3.7).

\begin{figure}
{\epsfxsize=8.truecm
\epsfbox[18 144 592 718]{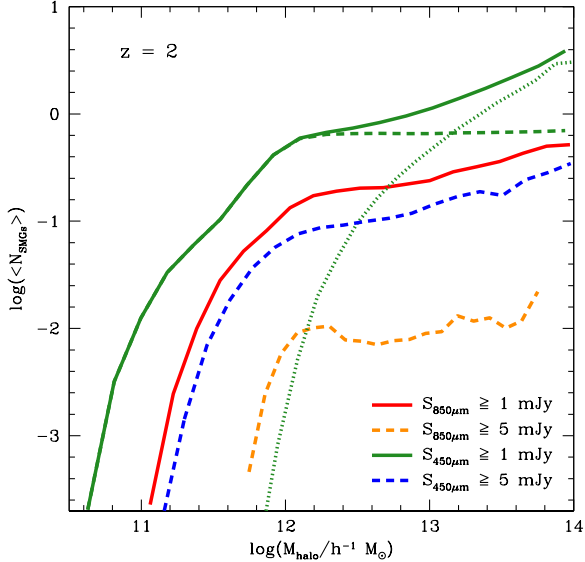}}
\caption{The predicted halo occupation distribution (HOD) 
of $z=2$ sub-mm selected galaxies.  
The dashed blue and orange lines show the HODs for galaxies 
brighter than 5~mJy at 450 and 850~$\mum$, while the green 
and red lines show the HODs for galaxies brighter than 1~mJy. 
The green dotted and dashed lines show the HOD of satellite 
and central galaxies with $S_{450\mu m}\ge 1$~mJy, respectively.}
\label{fig:clustering.hod}
\end{figure}

We gain further insight into the clustering predicted by the model by
plotting in Fig.~\ref{fig:clustering.hod} the mean number of sub-mm
selected galaxies as a function of the halo mass, generally refered to
as the halo occupation distribution or HOD \citep{benson00, cooray02,
berlind03}. For completeness, we compute the median halo mass of 
our samples: for SMGs with $S_{850\mu m}\ge 5$~mJy and 1~mJy, we find 
a median mass of $1.4 \times 10^{12} h^{-1}$ M$_{\odot}$ and 
$9.3 \times 10^{11} h^{-1}$ M$_{\odot}$, respectively; whereas for 
$S_{450\mu m}\ge 5$~mJy and 1~mJy selected galaxies, we determine 
$9.3 \times 10^{11} h^{-1}$ M$_{\odot}$ and $7.3 \times 10^{11} h^{-1}$ M$_{\odot}$.
Fig.~\ref{fig:clustering.hod} shows that both our $850\,\mu $m
selected samples display a mean number of galaxies below unity 
over most of the range of halo masses. For example, in our model 
we find on average one S$_{850\mu m}\ge 5$~mJy sub-mm galaxy for 
every $\sim 100$ dark matter haloes of mass 
$\sim 10^{13}\,h^{-1}$~M$_{\odot}$. This illustrates the need to 
consider a large number of halo merger histories in order 
to make robust predictions, a point we made previously for 
luminous red galaxies (Almeida et~al. 2008). 
We also predict that these massive haloes, with $M_{\rm halo} \ge
10^{13}\,h^{-1}$~M$_{\odot}$, will accommodate more than one SMG with
$S_{450\mu \rm m}\ge 1$~mJy (in our simulation, some haloes host more than
3 SMGs).  If this was not the case, the two-point
correlation function would tend to $\xi \sim -1$ on scales smaller
than the typical size of the host haloes \citep[see][]{benson00}. For
the case of $S_{450\mu m}\ge 1$~mJy we plot the contributions of
central and satellite galaxies to the HOD separately. The central 
galaxy HOD is seen to flatten above a certain mass, while the 
satellite HOD continues to rise and becomes a power law. 
It is interesting to note in this case that the HOD for central 
galaxies does not reach unity for any halo mass;  
around 70\% of haloes with masses $\ge 10^{12} h^{-1} M_{\odot}$ 
contain central galaxies brighter than 1 mJy at $450\,\mu$m.

\begin{figure}
{\epsfxsize=8.truecm
\epsfbox[18 144 592 718]{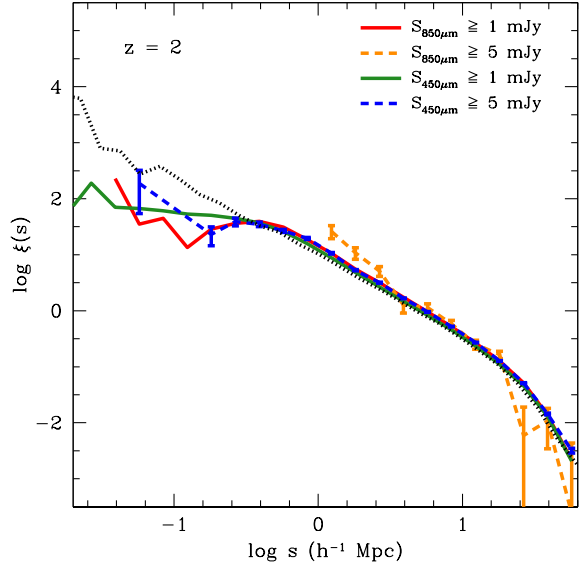}}
\caption{The redshift space two-point correlation function of sub-mm
  selected galaxies. The correlation function for $450\,\mu$m and
  $850\,\mu$m selected galaxies with flux densities brighter than 5
  mJy, are represented by the dashed blue and orange lines,
  respectively, while solid green and red lines show galaxies with
  fluxes S$_{450\mu m}\ge 1$ mJy and S$_{850\mu m}\ge 1$
  mJy. The errorbars show the $1\sigma$ Poisson errors derived from
  the number of pairs. For comparison we plot the S$_{450\mu m}\ge 1$
  mJy real-space correlation function, using a dotted black line.}
\label{fig:clustering.redshift.z2}
\end{figure}

\subsection{The redshift-space correlation function}

Galaxy surveys usually use redshift to infer the radial distance to a
galaxy, and the resulting measurement of clustering is said to be in
redshift space. We therefore need to take into account the
contribution of the peculiar velocity, induced by inhomogeneities in
the galaxy's surrounding density field, to the position of a galaxy
inferred from its redshift.  To model redshift space, we perturb the
position of the galaxy along one of the cartesian axes, $x$, by the
peculiar velocity of the galaxy along this axis, scaled by the
appropriate value of the Hubble parameter.  This corresponds to the
redshift position as viewed by a distant observer.  The redshift space
correlation function for galaxies selected by their sub-mm flux is
plotted in Fig.~\ref{fig:clustering.redshift.z2}.  The impact of
including the peculiar velocities on the correlation function
(redshift space distortions) depends on scale.  On intermediate and
large scales, the bulk motions of galaxies towards large scale
structures generate an amplification of the amplitude of the
correlation function \citep[see the comparison on large scales
in][]{jennings10}. On small scales, $r \lsim 3 \, h^{-1}$~Mpc, the
peculiar motions of galaxies within structures lead to a damping of
the correlation function. In this case there is an apparent stretching
of the structure in redshift space, which dilutes the number of SMGs
pairs.  As a consequence, if we calculate the correlation length in
redshift space, $s_0$, following the definition given by
Eq.~\ref{eq:r0}, i.e. $s_0$ is given by $\xi(s_0) = 1$, we obtain:
$s_0 = 6.4 \pm 0.5\,h^{-1}$Mpc for galaxies with $S_{850\mu m}\ge
5$~mJy, and $s_0 = 5.63 \pm 0.02 \,h^{-1}$Mpc for fainter galaxies
$S_{850\mu m}\ge 1$~mJy, slightly larger than the real-space values
given in \S\ref{sec:xi-real-space}.

\subsection{The evolution of the correlation length}
\label{section:z}

\begin{figure}
{\epsfxsize=8.truecm
\epsfbox[18 144 592 718]{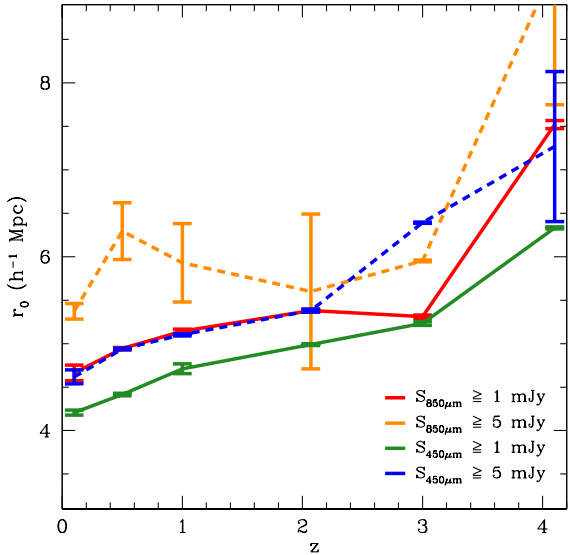}}
\caption{Evolution of the comoving correlation length with redshift
  for galaxies selected with $S_{850\mu m} \ge 1$ mJy (red line),
  $S_{850\mu m} \ge 5$ mJy (dashed orange line), $S_{450\mu m} \ge 1$
  mJy (solid green line), and $S_{450\mu m} \ge 5$ mJy (dashed blue
  line).  }
\label{fig:clustering.z}
\end{figure}

Having computed the spatial correlation functions of SMGs at $z=2$, it
is natural to study the evolution with redshift of the clustering
properties of these galaxies. In Fig.~\ref{fig:clustering.z} we plot
the evolution of the comoving real-space correlation length, $r_0$,
over the redshift interval $z=0-4$, for galaxies selected to be
brighter than either 1~mJy or 5~mJy at 450 or 850~$\mum$.  Here we
determine $r_0$ by finding the scale at which $\xi(r_0) = 1$.  The
wavelengths quoted are in the observer's frame, which means that, for
example, at $z=4$ $\lambda = 850\,\mu$m corresponds to a rest-frame
wavelength $\lambda_{\rm rest}\approx 170\,\mu$m.  Over the typical
redshifts at which SMGs are found (i.e. around $z=2$),
Fig.~\ref{fig:clustering.z} shows that the comoving correlation length
is approximately constant. There is an increase in correlation length
beyond $z\sim3$ and a decrease at $z<1$, but very few galaxies appear
in our samples at these redshifts.  Fig.~\ref{fig:clustering.z}
indicates that the correlation length of galaxies selected at $\lambda
= 450\,\mu$m is usually smaller than of $850\,\mu$m selected galaxies
at the same flux limit. This is not surprising due to the fact that
S$_{450\mu {\rm m}}/$S$_{850\mu {\rm m}} \approx 3$ (see the colour
distribution in Fig.~\ref{fig:colour}), i.e. the same galaxy would
appear about 3 times brighter at $\lambda = 450\,\mu$m than at
$\lambda = 850\,\mu$m.  This would then translate into a longer
comoving correlation length (as we will see in
Fig.~\ref{fig:clustering.sv}).

\subsection{Angular correlation function}

The simplest measure of clustering in photometric galaxy surveys 
is the angular two-point correlation function, $w(\theta)$, which 
is a weighted projection of $\xi(r)$ on the sky. The angular correlation 
function is used to measure clustering whenever redshift information 
is not available.  In this case, the probability of finding two
objects separated by angle $\theta$ is similar to Eq.~\ref{eq:prob1}:
\begin{equation}
 \label{eq:prob2}
 \delta P(\theta) = \bar{\eta}^2\,[1 + w(\theta)]\,\delta \Omega_1\,\delta \Omega_2,
\end{equation}
where $\bar{\eta}$ is the surface density of objects and $\delta \Omega_{\rm i}$ 
is an element of solid angle.

The galaxy formation predicts the spatial two-point correlation function, $\xi(r)$, 
and the redshift distribution of SMGs. From these predictions, it is straightforward 
to calculate the angular correlation function using Limber's equation \citep{limber53}. 
In a spatially flat universe, we have:
\begin{equation}
 w(\theta) = 2\,\frac{\int_0^\infty {\rm d}u \int_0^\infty x^4\,\Psi^2(x)\,
\xi(r,z)\,{\rm d}x} {\bigl[\int_0^\infty x^2\,\Psi(x)\,{\rm d}x\bigr]^2},
\end{equation}
where $u$ is related to the comoving distance $x$ by $r^2 = u^2 + x^2 - 2\,x u \cos\theta$.
The selection function, $\Psi(x)$, gives the probability that a sub-mm galaxy at a 
distance $x$ is detected in the survey, and is defined by:
\begin{equation}
\aleph = \int_0^\infty x^2\,\Psi(x)\,{\rm d}x = 
\frac{1}{\Omega_{\rm s}} \int_0^\infty N(z)\,{\rm d}z,
\end{equation}
where $\aleph$ is the surface density of galaxies, $\Omega_{\rm s}$ the solid angle covered by 
the survey, and $N(z)$ gives the number of sources within $z$ and $z + dz$.

\begin{figure}
{\epsfxsize=8.truecm
\epsfbox[18 144 592 718]{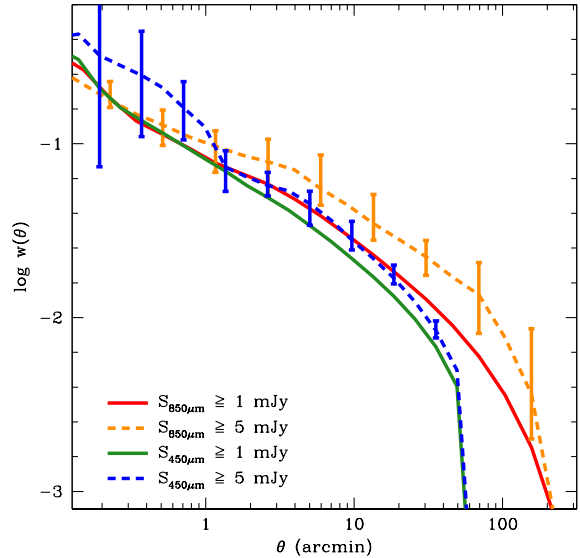}}
\caption{
  The two-point angular correlation function of sub-mm selected
  galaxies, computed directly from the spatial two-point correlation function.
  Galaxies selected at wavelengths of $450\,\mu$m and $850\,\mu$m 
  with fluxes brighter than 5~mJy are shown by dashed blue and orange 
  lines, while the green and red lines show galaxies with 
  $S_{450\mu {\rm m}}\ge 1$~mJy and $S_{850\mu {\rm m}}\ge 1$~mJy 
  respectively.  The errorbars show the $1\sigma$ Poisson errors 
  derived from the number of pairs in each bin. These are
  computed for an area of 1.3 $\deg^2$ at $450\,\mu$m and 35 $\deg^2$ at
  $850\,\mu$m, to match upcoming surveys, as described in the text. }
\label{fig:clustering.angular}
\end{figure}

In Fig.~\ref{fig:clustering.angular} we plot the angular correlation
function for $850\,\mu$m and $450\,\mu$m selected galaxies. We show
predictions for galaxies brighter than 1 mJy and 5 mJy. We also show 
errorbars derived from the number of pairs in each bin in angular 
separation. These are computed for areas of 1.3$\deg^2$ and 35$\deg^2$ 
at $450\,\mu$m and $850\,\mu$m respectively, chosen to match the survey 
areas planned with SCUBA-2.

Fig.~\ref{fig:clustering.angular} shows that between $1$ and $10$ 
arcminutes , $w(\theta)$ can be approximated by a power law:
\begin{equation}
\label{eq:wtheta}
w(\theta) = \biggl(\frac{\theta}{\theta_0}\biggr)^{1+\gamma},
\end{equation}
where $\gamma$ is the same as in Eq.~\ref{eq:xi}. The amplitude of
clustering is characterised by $\theta_0$. This scale can vary from
one sample to another because of differences in the intrinsic
clustering and the redshift distribution of galaxies. Chance
projections of galaxy pairs, as would occur more often in a catalogue
which spanned a broad redshift interval, lead to a dilution of the
clustering signal, which results in $\theta_0$ decreasing with
increasing survey depth \citep{peebles80}.  The predicted angular
correlation function steepens slightly below $\approx 1 $arcmin.
Fig.~\ref{fig:clustering.angular} shows that bright SMGs are more
clustered than faint SMGs.  Fitting Eq.~\ref{eq:wtheta} to our samples
over the range of angular scales $1 < \theta < 10$ arcmin, with a
fixed $\gamma = -1.7$ (see previous section), we find a clustering
scale of 
$\theta_0(S_{850\mu m} \ge 5$ mJy$) = 0.028 \pm 0.012$~arcmin. 
For galaxies selected by $\lambda = 450\,\mu$m flux, we find
$\theta_0(S_{450\mu m}\ge 5$ mJy$) = 0.022 \pm 0.007$~arcmin.

Observational estimates of the angular correlation function of SMGs
suggest that these galaxies are strongly clustered. However, a
consensus on the clustering amplitude of SMGs is yet to be reached.
For example, using a small sample of SMGs observed with the SCUBA
camera, \citet{scott02} found evidence of strong clustering on scales
of 1 -- 2 arcmin. In a follow-up study, \citet{scott06} re-reduced and
combined different SCUBA surveys to measure the angular clustering and
inferred a clustering scale $\theta_0 \approx 0.6-0.8$ arcmin,
depending on the precise choice of flux limit and signal-to-noise
cut-off used to construct maps of SMGs. The challenge of measuring the
clustering signal with such small samples is apparent from the size of
the integral constraint correction typically applied.  The integral
constraint takes into account the fact that the true mean density of
SMGs is not known when estimating fluctuations in the galaxy
distribution. Instead, the density of the sample itself is used as a
substitute for the true mean density. If the sample density is
different from the true mean then the fluctuation level is
misestimated, leading to a bias in $w(\theta)$. This effect is
important for small samples and is even more severe if the galaxies
are also strongly clustered.  Scott et~al. applied a correction of a
factor of $\approx 2$ to their measured clustering to account for the
integral constraint.  A subsequent measurement of clustering in a
sample constructed with the LABOCA detector by Weiss et~al. (2009)
gives a correlation angle of $\theta_{0} = 0.23 \pm 0.12$ arcmin,
which is substantially weaker that the Scott et~al. result. The Weiss
et~al. estimate is still much larger than our model prediction, but
differs from it by only 1.8$\sigma$.  Blain et~al. (2004) made an
indirect measurement of the spatial correlation function using radial
pair counts of SMGs with spectroscopic redshifts, and obtained $r_{0}
= 6.9 \pm 2.1 h^{-1}\,$Mpc. For spatial clustering that is fixed in
comoving coordinates, and using the observed redshift distribution of
SMGs (Chapman et~al. 2004), the Blain et~al. result implies $\theta_0
= 0.04 \pm 0.01$ arcmin, which is comparable to our model prediction
However, the method used by Blain et~al. does not take into account
fluctuations in the sample density, and should perhaps be viewed as
providing a lower limit on the clustering strength.

\section{The dependence of clustering strength on galaxy properties}

\begin{figure}
{\epsfxsize=8.truecm
\epsfbox[18 144 592 718]{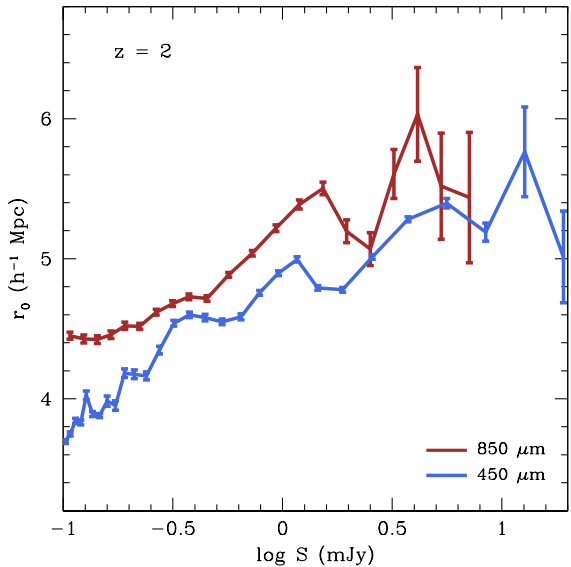}}
\caption{
The dependence of the comoving correlation length, $r_{0}$, on sub-mm flux 
for galaxies at $z=2$. The predictions for SMGs selected by 
their emission at $850\,\mu$m are shown by the red line, 
and those for $450\,\mu$m selected sub-mm galaxies are
plotted in blue. 
The flux plotted on the x-axis is correspondingly at either 
850 or 450~$\mum$.
}
\label{fig:clustering.sv}
\end{figure}

Here we consider the dependence of clustering strength on galaxy 
properties. We present the predictions for the two-point correlation 
function as a function of sub-mm flux, halo and stellar mass, 
S$_{450\mu {\rm m}}$/S$_{850\mu {\rm m}}$ colour and the quiescent 
or starburst nature of galaxies at $z=2$.

\subsection{Dependence of clustering on sub-mm flux}

In Fig.~\ref{fig:clustering.sv} we plot the comoving
correlation length, $r_0$ (computed using Eq.~6), 
as a function of sub-mm flux, S$_{\nu}$, for
galaxies at $z=2$.  The plot shows that brighter galaxies 
have larger correlation lengths (see also Figs.~\ref{fig:clustering.real.z2} 
and ~\ref{fig:clustering.angular}). However, the dependence of clustering 
strength on luminosity is fairly weak, with a change of 50\% in correlation 
length on changing flux by a factor of a hundred. This behaviour can be 
understood as a consequence of brighter galaxies being found predominately 
in more massive haloes, for which the bias is greater than unity and increases 
strongly with mass (Angulo et~al. 2009). 
As already pointed out in relation to Fig.~\ref{fig:clustering.real.z2}, 
for the same flux limit, $450\,\mu $m selected galaxies are less clustered 
than their $850\,\mu $m counterparts, with correlation lengths that 
are typically smaller by $\Delta r_0 \approx 0.4 \,h^{-1}$Mpc. 
This difference remains roughly constant throughout the range of sub-mm 
fluxes explored (in Fig.~\ref{fig:clustering.colour} we will see that 
this is a consequence of the fact that the median 
S$_{450\mu m}$/$S_{850\mu m}$ colour is approximately $\approx 3$).

\begin{figure}
{\epsfxsize=8.truecm
\epsfbox[18 144 592 718]{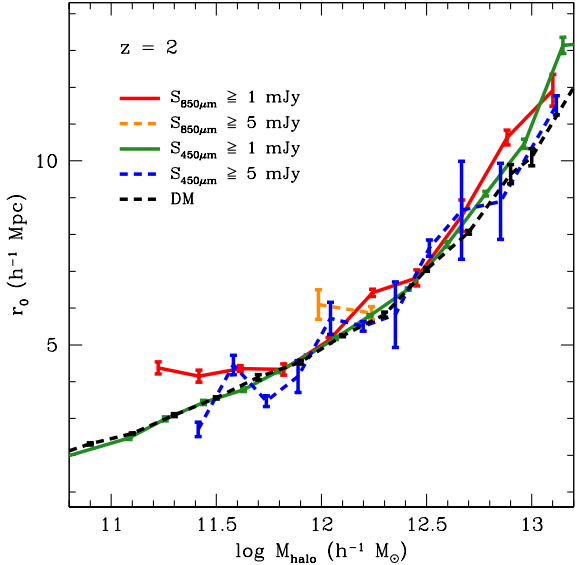}}
\caption{The dependence of the comoving correlation length, $r_{0}$,
on host halo mass, $M_{\rm halo}$, for galaxies at $z=2$ (for
differential bins in halo mass).  The red and green solid lines
display the relation predicted for galaxies with $S_{850\mu m} \ge 1$
mJy and $S_{450\mu m} \ge 1$ mJy respectively. Brighter sub-mm
galaxies, with fluxes above 5 mJy, are shown by the dashed orange
($850\,\mu$m) and dashed blue ($450\,\mu$m) lines. The dashed black
line shows the dependence of correlation length on halo mass for all
haloes (regardless of whether they host an SMG or not).  }
\label{fig:clustering.mhalo}
\end{figure}

\begin{figure}
{\epsfxsize=8.truecm
\epsfbox[18 144 592 718]{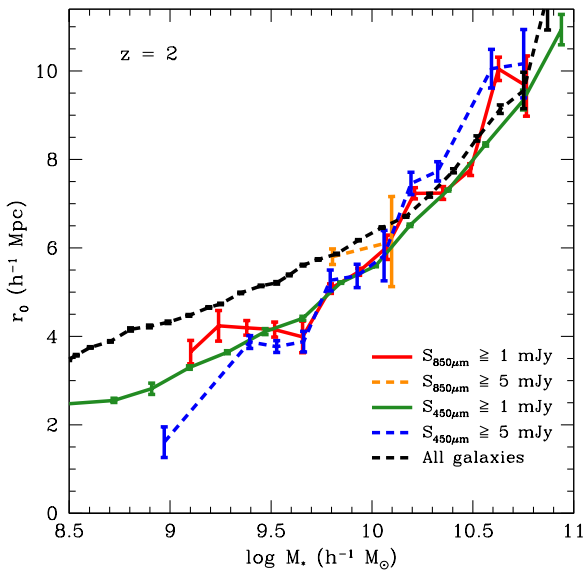}}
\caption{
The dependence of the comoving correlation length, $r_{0}$, 
on total stellar mass, $M_*$, for galaxies at $z=2$. The red 
and green solid lines display the predictions for galaxies 
with $S_{850\mu m} \ge 1$ mJy and $S_{450\mu m} \ge 1$ mJy, respectively. 
Brighter sub-mm galaxies, with fluxes brighter than 5 mJy, are 
shown by the dashed orange ($850\,\mu$m) and dashed blue 
($450\,\mu$m) lines. The relation between $r_0$ and stellar mass 
for all the galaxies in the simulation, regardless of their 
sub-mm flux, is shown by the dashed-black line.
}
\label{fig:clustering.mstar}
\end{figure}

\subsection{Dependence of clustering strength on halo and stellar mass}

Fig.~\ref{fig:clustering.mhalo} shows the dependence of $r_{0}$ on the
mass of the dark matter halo which hosts the SMG, for discrete bins in
halo mass (the bin width is adjusted to contain a representative
sample of galaxies). Remarkably, there is little difference between
the clustering strength predicted for bright and faint samples of SMGs
within a given bin of host halo mass.  Moreover, the SMG clustering is
essentially the same as that of all of the dark matter haloes in the
mass bin, regardless of whether or not they contain an SMG. This means
that the clustering strength of SMGs is driven purely by the mass of
the host halo, and shows little or no dependence on a second property
of the halo, such as its formation time or spin. One might have
expected to see a dependence of the clustering strength at a given
halo mass on SMG flux if this selection favoured host haloes which
had, for example, formed more recently than the overall population at
a given mass \citep[see, e.g.][]{gao05, percival03}.

We now consider the dependence of the correlation length $r_{0}$ on 
stellar mass, $M_*$, which we plot in Fig.~\ref{fig:clustering.mstar}. 
Following earlier plots, we display the relation for SMGs with 
fluxes brighter than $1$ mJy at $450 \mu$m and $850 \mu$m, and 
for brighter samples with fluxes $\ge 5$ mJy at both wavelengths. 
At $z=2$, the median stellar masses of the model samples are 
$M_* = 9.9\times 10^{9}\,h^{-1}$M$_{\odot}$ for galaxies with 
$S_{850} \ge 1 $mJy, and $M_* = 9.7\times 10^{9}\,h^{-1}$M$_{\odot}$ for 
those with $S_{450} \ge 1 $mJy \citep[see also][]{gonzalez10}. 
Comparisons with observational estimates of stellar masses are  
of limited use, as the results depend critically on the assumption 
made about the form of the stellar IMF (see Lacey et~al. 2010a for an 
expanded discussion). For example, \citet{hainline10} estimated the 
stellar masses of $\sim 70$ SMGs with $S_{850\mu m} \approx 5$ mJy 
and found a median stellar mass of $7 \times 10^{10} h^{-1}$M$_{\odot}$, 
when using a Kroupa (2001) IMF. At the same flux limit, our model 
predicts a median stellar mass of $2 \times 10^{10} h^{-1}$M$_{\odot}$.

There is a tight, monotonic correlation between the comoving
correlation length, $r_{0}$, and stellar mass, $M_*$ (Fig.~\ref{fig:clustering.mstar}). 
Massive galaxies are more strongly clustered than galaxies with lower 
stellar masses, implying that more massive galaxies tend to be hosted by 
more massive haloes. Interestingly, all of our samples show a similar 
correlation length for a given total stellar mass, i.e. the relation between
correlation length and total stellar mass does not depend on the sub-mm 
luminosity of the galaxy. We find that the correlation length $r_{0}$ varies 
roughly linearly with $\log M_*$ in the range $\log M_* \approx [9.5, 10.5] h^{-1}$M$_{\odot}$.  
Another important aspect is the fact that SMGs have, typically, weaker clustering 
than other galaxies of the same stellar mass. This is particularly evident 
for stellar masses below $10^{10} h^{-1}$M$_{\odot}$. The phenomenon is 
essentially due to the combination of two factors. Firstly, there is a 
difference in the host halo masses for satellite and central galaxies of 
the same stellar mass: at a given stellar mass, satellite galaxies inhabit
haloes which are roughly ten times more massive than central galaxies. 
Secondly, SMGs galaxies with stellar masses lower than $10^{10} h^{-1}$M$_{\odot}$
are mainly central galaxies, whereas the fraction of satellite galaxies for 
the full galaxy population is approximately 30 per cent. Submillimetre 
galaxies with stellar masses M$_{*} \ge 10^{10} h^{-1}$M$_{\odot}$ 
display a similar fraction of satellite galaxies to that in the full sample. 
Currently, there are no observational determinations of this relation at 
the redshifts of interest for SMGs.

\begin{figure}
{\epsfxsize=8.truecm
\epsfbox[18 144 592 718]{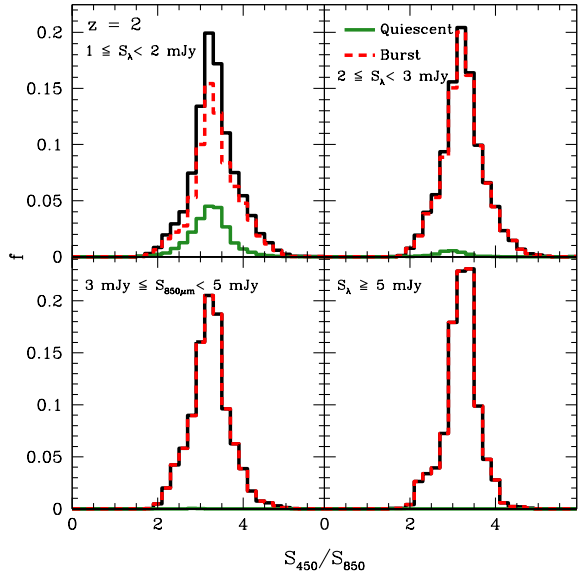}}
\caption{ Distribution of the S$_{450\mu {\rm m}}$/$S_{850\mu {\rm
m}}$ flux ratio or colour for SMGs at $z=2$, as a function of
850~$\mu$m flux density.  The solid red and dashed orange histograms
show the distribution for quiescent and starburst galaxies,
respectively, while the black line displays the combined
distribution. Quiescent galaxies only make a significant contribution
at the faintest fluxes (top left). The distributions are normalized to
integrate to unity.}
\label{fig:colour}
\end{figure}

\subsection{Dependence of clustering on sub-mm colour}

Before studying the dependence of the correlation length, $r_0$, on
sub-mm colour, as given by the S$_{450\mu {\rm m}}$/S$_{850\mu {\rm
m}}$ flux ratio, it is informative to first plot the colour
distribution itself, which we show in Fig.~\ref{fig:colour}.  Here, we
separate galaxies into bins of S$_{850\mu m}$ flux and discriminate
between quiescent and starburst galaxies. In our model, the S$_{450\mu
{\rm m}}$/$S_{850\mu {\rm m}}$ colour distribution of $\lambda =
850\,\mu$m selected galaxies peaks around $\approx 3.2$.  Furthermore,
we find that the mode and shape of the colour distribution does not
change significantly with $S_{850\mu {\rm m}}$ flux, i.e.  the
colour--luminosity relation of sub-mm galaxies with $S_{850\mu {\rm
m}} \ge 1$ mJy is roughly constant. Also, there is very little
difference between the colour distribution of starburst and quiescent
galaxies (note, however, that, at lower fluxes, the fraction of sub-mm
galaxies which are forming stars quiescently increases).

\begin{figure}
{\epsfxsize=8.truecm
\epsfbox[18 144 592 718]{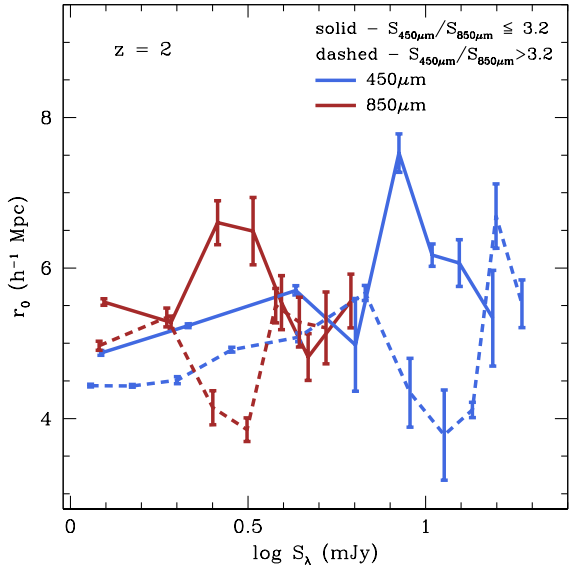}}
\caption{ Dependence of the correlation length, $r_{0}$, on $S_{450\mu
{\rm m}}/S_{850\mu {\rm m}}$ colour for galaxies at $z=2$. SMGs
selected by their emission at $\lambda = 850\,\mu$m are represented by
the red line, while the predictions for galaxies selected at $\lambda
= 450\,\mu$m are plotted in blue. Solid and dashed lines show the
correlation length for galaxies split by sub-mm colour, $S_{450\mu
{\rm m}}/S_{850\mu {\rm m}}$, as indicated by the key.  }
\label{fig:clustering.colour}
\end{figure}

The observed distribution of $450\mu$m/$850\mu$m colours of SMGs has
not yet been accurately measured, but the $350\mu$m/$850\mu$m colours
of SMGs in the same model were investigated by \citet{swinbank}, who
found the predicted colours to be in agreement with observed
values. They also found that the median $350\mu$m/$850\mu$m colour of
model SMGs could be fit by a modified blackbody spectrum $L_{\nu}
\propto B_{\nu}(T)\, \nu^{\beta}$, with $\beta=1.5$ and an effective
dust temperature of $T=32$~K. This modified blackbody implies a colour
$S_{450\mu m}/S_{850\mu m}\approx 3.5$ for SMGs at $z=2$, which agrees
well with the distribution plotted in
Fig.~\ref{fig:clustering.colour}.

In Fig.~\ref{fig:clustering.colour} we plot the correlation length,
$r_0$, as a function sub-mm flux for two samples split at a sub-mm
colour of S$_{450\mu m}$/S$_{850\mu m}= 3.2$.  The dependence of $r_0$
on luminosity is similar to that shown in
Fig.~\ref{fig:clustering.sv}.  The figure also hints that redder
galaxies, i.e. those with S$_{450\mu m}$/S$_{850\mu m} \le 3.2$, are
more clustered than bluer galaxies with S$_{450\mu m}$/S$_{850\mu m}
\ge 3.2$ (for both $\lambda = 850\,\mu$m and $\lambda = 450\,\mu$m
selected galaxies). For example, for S$_{850\mu m} \approx 3$ mJy
galaxies, $r_0$ can differ by a factor of $\approx 1.6$ between the
red and blue samples.

\begin{figure}
{\epsfxsize=8.truecm
\epsfbox[18 144 592 718]{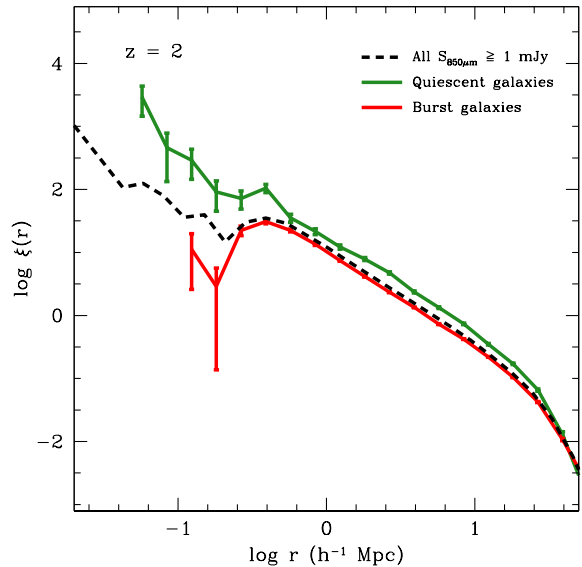}}
\caption{The real-space two-point correlation function of 
quiescent (red line) and starburst (orange) sub-mm galaxies, 
selected by S$_{850\mu m}\ge 1$ mJy at $z = 2$. The dashed 
black line shows the correlation function for all galaxies 
with fluxes brighter than 1 mJy.}
\label{fig:xi.quiescentburst}
\end{figure}

Having plotted the colour distributions of starburst and quiescent
galaxies and the relation between $r_0$ and sub-mm colour, it is
useful to study the clustering of quiescent and starburst galaxies
separately. As mentioned in connection with Fig.~\ref{fig:colour},
quiescent galaxies only make a significant contribution at lower
fluxes, $S_{850\mu m} \le 2$ mJy. In Fig.~\ref{fig:xi.quiescentburst},
we plot the two-point correlation function in real-space of sub-mm
galaxies selected by S$_{850\mu m} \ge 1$ mJy at $z = 2$. We also plot
the correlation functions of quiescent and starburst galaxies
separately.  
Fig.~\ref{fig:clustering.colour} reveals that, in our model, quiescent
galaxies are more clustered than burst galaxies of similar sub-mm
luminosities, on all scales. This is mostly due to the fact that
quiescent SMGs are more massive than burst SMGs, due to both the
top-heavy IMF and shorter star formation timescale in bursts. We
compute $r_0 = 5.2\,h^{-1}$ Mpc for burst galaxies with $S_{850\mu m}
\ge 1$ mJy, and $r_0 = 7.2\,h^{-1}$ Mpc for quiescent galaxies
brighter than the same flux limit.

\section{Discussion and Conclusions}
\label{section:conclusions}

In this paper we have applied the technique introduced by \citet{ann},
based on artificial neural networks, to predict the spatial and
angular clustering of sub-mm selected galaxies (SMGs) in a
$\Lambda$CDM universe. The ANN allows us to rapidly mimic the
predictions of a hybrid code, made up of the \g~semi-analytical model
of galaxy formation, which predicts the full star formation and merger
histories of galaxies, and the \gr~code, which computes the spectral
energy distributions of galaxies self-consistently.  This makes it
possible for large N-body simulations of the hierarchical growth of
structure in the dark matter to be populated with galaxies with full
SED coverage from the UV through to the radio.  The ANN technqiue
requires as input a small set of galaxy properties predicted by \g.
We use the algorithm to populate the Millennium Simulation with
$\lambda =850\,\mu$m and $\lambda = 450\,\mu$m selected galaxies, with
$S_{\lambda} \ge 1$ mJy.  The accuracy of the artificial neural
network is notable: for SMGs at $z=2$, we are able to reproduce the
luminosity of $\approx 95$ per cent of galaxies to within 10 per cent
of the true values. The errors introduced by this technique in the
determination of clustering properties are negligible.

We have presented predictions for the two-point spatial and angular
correlation functions, for samples of galaxies selected at $850\,\mu$m
and $450\,\mu$m.  At $z=2$, we predict a comoving correlation length
of $r_0 = 5.6\pm 0.9\,h^{-1}$Mpc for galaxies with $S_{850\mu m} \ge
5$ mJy and $r_0 = 5.38\pm 0.02\,h^{-1}$Mpc for fainter galaxies,
brighter than 1 mJy. The former value is in good agreement with the
indirect observational estimate by \citet{blain04} who found $r_0 =
6.9\pm 2.1\,h^{-1}$Mpc for a small sample of SMGs with $S_{850\mu m}
\sim 5$~mJy. Galaxies selected at $\lambda = 450\,\mu$m are less
clustered than those identified at $850\,\mu$m galaxies, to the same
flux limit.

Not surprisingly, the correlation length of sub-mm selected galaxies
evolves with redshift: galaxies selected at higher redshifts display
larger comoving correlation lengths, i.e. with increasing redshift
they trace more overdense (and consequently, rarer) regions of the
Universe. This behaviour is similar to that expected for the growth of
structures in a $\Lambda$CDM universe, and further supports the idea
that sub-mm galaxies are biased tracers of the underlying mass.

We have also studied the dependence of clustering on some properties
of our sub-mm galaxies. Generally, we found a strong dependence of
clustering on luminosity, with brighter galaxies displaying higher
correlation lengths. We predict a tight correlation
between halo mass and clustering scale, with sub-mm galaxies hosted by
massive haloes being more clustered than those hosted by less massive
ones. We find that the relation found for SMGs is not significantly
different from the clustering of all haloes of the same mass. A
similar behaviour is observed for the correlation between clustering
and total stellar mass. We find that more massive sub-mm galaxies are
more clustered than less massive one (in fact, massive galaxies are
predominately found in more massive haloes -- haloes with higher
effective bias): in our model, $r_0$ changes by a factor of $\approx
2$ within the range $\log M_* \approx [9.5, 10.5]\,h^{-1}$M$_{\odot}$. 
The dependence of clustering on sub-mm colour, $S_{450\mu m}/S_{850\mu
m}$, is not so clear.  We find a weak positive correlation between
$r_0$ and colour, with redder galaxies being more strongly
clustered. Finally, we predict that for galaxies selected to have
$S_{850\mu m} \ge 1$~mJy, quiescent galaxies are more clustered than
those undergoing a burst of star formation.

It is important to note that current observational measurements of the
clustering properties of sub-mm selected galaxies are limited due to
poor statistics.  At $z\sim 2$, estimations of the correlation
function currently rely on samples with fewer than $\sim 100$
galaxies. However, this picture will soon be improved. A more detailed
comparison will soon be possible with the forthcoming instruments,
such as the SCUBA-2, which will allow a much deeper and wider survey
of the submillimetre population up to redshift 4 than current
instruments.  A more accurate determination of the clustering will, by
then, impose serious constrains on the theories of galaxy formation
and evolution.

\subsection*{ACKNOWLEDGEMENTS}
C. A. gratefully acknowledges support from the NSFC Research Fellowship 
for International Young Scientists, grant no 10950110319. This work was 
supported by the Chinese Academy of Sciences, grant no. 2009YB1, and 
by the Science and Technology Facilities Council rolling grant to the ICC.


\begin{thebibliography}{99}
{\small
\bibitem[\protect\citeauthoryear{Alexander et al.}{2003}]{alexander03}
Alexander D. M., Bauer F. E., Brandt W. N., Hornschemeier A. E., Vignali C., Garmire G. P.,
Schneider D. P., Chartas G., Gallagher S. C., 2003, MNRAS, AJ, 125, 383
\bibitem[\protect\citeauthoryear{Alexander et al.}{2005}]{alexander05}
Alexander D. M., Bauer F. E., Chapman S. C., Smail I., Blain A. W., Brandt W. N., Ivison R. J.,
2005, ApJ, 632, 736
\bibitem[\protect\citeauthoryear{Almeida et al.}{2007}]{almeida07}
Almeida C., Baugh C. M., Lacey C. G., 2007, MNRAS, 376, 1711
\bibitem[\protect\citeauthoryear{Almeida et al.}{2008}]{almeida08}
Almeida C., Baugh C. M., Wake D. A., Lacey C. G., Benson A. J., Bower R. G., Pimbblet K.,
2008, MNRAS, 386, 2145
\bibitem[\protect\citeauthoryear{Almeida et al.}{2010}]{ann}
Almeida C., Baugh C. M., Lacey C. G., Frenk C. S., Granato G. L., Silva L., Bressan A.,
2010, MNRAS, 402, 544
\bibitem[\protect\citeauthoryear{Angulo et al.}{2008}]{angulo08}
Angulo R. E., Baugh C. M., Frenk C. S., Lacey C. G., 2008, MNRAS, 383, 755
\bibitem[\protect\citeauthoryear{Angulo et al.}{2009}]{angulo09}
Angulo R. E., Lacey C. G., Baugh C. M., Frenk C. S., 2009, MNRAS, 399, 983
\bibitem[\protect\citeauthoryear{Barger et al.}{1998}]{barger98}
Barger A. J., Cowie L. L., Sanders D. B., Fulton E., Taniguchi Y., Sato Y.,
Kawara K., Okuda, H., 1998, Nat, 394, 248
\bibitem[\protect\citeauthoryear{Baugh et al.}{1999}]{baugh99}
Baugh C. M., Benson A. J., Cole S., Frenk C. S., Lacey C. G., 1999, MNRAS, 305, 21
\bibitem[\protect\citeauthoryear{Baugh et al.}{2005}]{baugh05}
Baugh C. M., Lacey C. G., Frenk C. S., Granato G. L., Silva L., Bressan A., Benson A. J., Cole S.,
2005, MNRAS, 356, 1191
\bibitem[\protect\citeauthoryear{Baugh}{2006}]{baugh06}
Baugh C. M., 2006, Rep. Prog. Phys., 69, 3101
\bibitem[\protect\citeauthoryear{Benson et al.}{2000}]{benson00}
Benson A. J., Baugh C. M., Cole S., Frenk C. S., Lacey C. G., 2000, MNRAS, 316, 107
\bibitem[\protect\citeauthoryear{Benson et al.}{2003}]{benson03}
Benson A. J., Bower R. G., Frenk C. S., Lacey C. G., Baugh C. M., Cole S., 2003, ApJ, 599, 38
\bibitem[\protect\citeauthoryear{Benson}{2010}]{benson10}
Benson A. J., 2010, Phys. Rep., in press, (arXiv:1006.5394)
\bibitem[\protect\citeauthoryear{Berlind \& Weinberg}{2002}]{berlind02}
Berlind A. A., Weinberg D. H., 2002, ApJ, 575, 587
\bibitem[\protect\citeauthoryear{Berlind et al.}{2003}]{berlind03}
Berlind A. A., et al., 2003, ApJ, 539, 1
\bibitem[\protect\citeauthoryear{Biggs \& Ivison}{2008}]{biggs08}
Biggs A. D., Ivison R. J., 2008, MNRAS, 385, 893
\bibitem[\protect\citeauthoryear{Blain et al.}{2002}]{blain02}
Blain A. W., Smail I., Ivison R. J., Kneib J. -P., Frayer D. T., 2002, Phys. Rep., 369, 111
\bibitem[\protect\citeauthoryear{Blain et al.}{2004}]{blain04}
Blain A. W., Chapman S. C., Smail I., Ivison R., 2004, ApJ, 611, 725
\bibitem[\protect\citeauthoryear{Borys et al.}{2005}]{borys05}
Borys C., Smail I., Chapman S. C., Blain A. W., Alexander D. M., Ivison R. J.,
2005, ApJ, 635, 853
\bibitem[\protect\citeauthoryear{Bower et al.}{2006}]{bower06}
Bower R. G., Benson A. J., Malbon R., Helly J. C., Frenk C. S., Baugh C. M., Cole S., Lacey C. G.,
2006, MNRAS, 370, 645
\bibitem[\protect\citeauthoryear{Bressan et al.}{2002}]{bressan02}
Bressan A., Silva L., Granato G. L., 2002, A\&A, 392, 377
\bibitem[\protect\citeauthoryear{Brodwin et al.}{2008}]{brodwin08}
Brodwin M., et al., 2008, ApJ, 687, 65
\bibitem[\protect\citeauthoryear{Bruzual \& Charlot}{2003}]{bruzual03}
Bruzual G., Charlot S., 2003, MNRAS, 344, 1000
\bibitem[\protect\citeauthoryear{Casey et al.}{2009}]{casey09}
Casey C. M., Chapman S. C., Muxlow T. W. B., Beswick R. J., Alexander D. M.,
Conselice C. J., 2009, MNRAS, 395, 1249
\bibitem[\protect\citeauthoryear{Chapman et al.}{2000}]{chapman00}
Chapman et al., 2000, MNRAS, 319, 318
\bibitem[\protect\citeauthoryear{Chapman et al.}{2005}]{chapman05}
Chapman S. C., Blain A. W., Smail I., 2005, ApJ, 622, 772
\bibitem[\protect\citeauthoryear{Cole et al.}{2000}]{cole00}
Cole S., Lacey C. G., Baugh C. M., Frenk C. S., 2000, MNRAS, 319, 168
\bibitem[\protect\citeauthoryear{Cooray \& Sheth}{2002}]{cooray02}
Cooray A., Sheth R., 2002, Physics Reports, 372, 1
\bibitem[\protect\citeauthoryear{Cooray et al.}{2010}]{cooray10}
Cooray A., et al., 2010, A\&A, 518, 22
\bibitem[\protect\citeauthoryear{Croton et al.}{2006}]{croton06}
Croton D. J., et al., 2006, MNRAS, 365, 11
\bibitem[\protect\citeauthoryear{Dunne et al.}{2003}]{dunne03}
Dunne L., Eales S., Edmunds M., 2003, MNRAS, 341, 589
\bibitem[\protect\citeauthoryear{Eke et al.}{2004}]{eke04}
Eke V. R., et al., 2004, MNRAS, 355, 769
\bibitem[\protect\citeauthoryear{Elmegreen}{2009}]{elmegreen}
Elmegreen B. G., 2009, in The Evolving ISM in the Milky Way and Nearby Galaxies
\bibitem[\protect\citeauthoryear{Engelbracht et al.}{2006}]{engelbracht06}
Engelbracht C. W., et al., 2006, ApJ, 642, 127
\bibitem[\protect\citeauthoryear{Farrah et al.}{2006}]{farrah06}
Farrah D., et al., ApJ, 641, L17
\bibitem[\protect\citeauthoryear{Ferrara et al.}{1999}]{ferrara99}
Ferrara A., Bianchi S., Cimatti A., Giovanardi C., 1999, ApJS, 123, 423
\bibitem[\protect\citeauthoryear{Font et al.}{2008}]{font08}
Font A. S., et al., 2008, MNRAS, 389, 1619
\bibitem[\protect\citeauthoryear{Gao et al.}{2005}]{gao05}
Gao L., Springel V., White S. D. M., 2005, MNRAS, 363, L66
\bibitem[\protect\citeauthoryear{Gonz\'alez et al.}{2009}]{gonzalez09}
Gonz\'alez J. E., Lacey C. G., Baugh C. M., Frenk C. S., Benson A. J.,
2009, MNRAS, 397, 1254
\bibitem[\protect\citeauthoryear{Gonz\'alez et al.}{2010a}]{gonzalez10a}
Gonz\'alez J. E., Lacey C. G., Baugh C. M., Frenk C. S.,
2010a, MNRAS, submitted
\bibitem[\protect\citeauthoryear{Gonz\'alez et al.}{2010b}]{gonzalez10}
Gonz\'alez J. E., Lacey C. G., Baugh C. M., Frenk C. S.,
2010b, MNRAS, submitted (arXiv:1006.0230)
\bibitem[\protect\citeauthoryear{Gonzalez-Perez et al.}{2009}]{violeta}
Gonzalez-Perez V., Baugh C. M., Lacey C. G., Almeida C., 2009, MNRAS, 398, 497
\bibitem[\protect\citeauthoryear{Granato et al.}{2000}]{granato00}
Granato G. L., Lacey C. G., Silva L., Bressan A., Baugh C. M., Cole S., Frenk C. S.,
2000, ApJ, 542, 710
\bibitem[\protect\citeauthoryear{Guiderdoni et al.}{1998}]{guiderdoni98}
Guiderdoni B., Hivon E., Bouchet F. R., Maffei B., 1998, MNRAS, 295, 877
\bibitem[\protect\citeauthoryear{Guo et al.}{2010}]{guo10}
Guo, et al., 2010, MNRAS, submitted
\bibitem[\protect\citeauthoryear{Harker et al.}{2006}]{harker06}
Harker G., Cole S., Helly J., Frenk C., Jenkins A., 2006, MNRAS, 367, 1039
\bibitem[\protect\citeauthoryear{Hainline et al.}{2010}]{hainline10}
Hainline L. J., Blain A. W., Smail I., Alexander D. M., Armus L.,
Chapman S. C., Ivison R. J., 2010, MNRAS, submitted (arXiv:1006.0238)
\bibitem[\protect\citeauthoryear{Hughes et al.}{1997}]{hughes97}
Hughes D. H., Dunlop J. S., Rawlings S., 1997, MNRAS, 289, 766
\bibitem[\protect\citeauthoryear{Ivison et al.}{2000}]{ivison00}
Ivison R. J., Smail I., Barger A. J., Kneib J. -P., Blain A. W., Owen F. N.,
Kerr T. H., Cowie L. L., 2000, MNRAS, 315, 209
\bibitem[\protect\citeauthoryear{Jennings et al.}{2010}]{jennings10}
Jennings E., Baugh C. M., Pascoli S., 2010, MNRAS, submitted (arXiv:1003.4282)
\bibitem[\protect\citeauthoryear{Kaiser}{1987}]{kaiser87}
Kaiser N., 1987, MNRAS, 227, 1
\bibitem[\protect\citeauthoryear{Kennicutt}{1983}]{kennicutt}
Kennicutt R. C., 1983, ApJ, 272, 54
\bibitem[\protect\citeauthoryear{Kim et al.}{2009}]{kim09}
Kim H.-S., Baugh C. M., Cole S., Frenk C. S., Benson A. J., 2009, MNRAS, 400, 1527
\bibitem[\protect\citeauthoryear{Kim et al.}{2010}]{kim10}
Kim H-S., Baugh C. M., Benson A. J., Cole S., Frenk C. S., 
Lacey C. G., Power C., Schneider M., 2010, MNRAS, submitted (arXiv:1003.0008)
\bibitem[\protect\citeauthoryear{Kroupa}{2001}]{kroupa01}
Kroupa P., 2001, MNRAS, 322, 231
\bibitem[\protect\citeauthoryear{Lacey et al.}{2008}]{lacey08}
Lacey C. G., Baugh C. M., Frenk C. S., Silva L., Granato G. L., Bressan A.,
2008, MNRAS, 385, 1155
\bibitem[\protect\citeauthoryear{Lacey et al.}{2010a}]{lacey10a}
Lacey C. G., Baugh C. M., Frenk C. S., Benson A. J., Orsi A., Silva L., Granato G. L., 
Bressan, A.,2010a, MNRAS, 405, 2
\bibitem[\protect\citeauthoryear{Lacey et al.}{2010b}]{lacey10b}
Lacey C. G., Baugh C. M., Frenk C. S., Benson A. J., 
2010b, MNRAS, submitted (arXiv:1004.3545)
\bibitem[\protect\citeauthoryear{Lagos et al.}{2008}]{lagos08}
Lagos C. D. P., Cora S. A., Padilla N. D., 2008, MNRAS, 388, 587
\bibitem[\protect\citeauthoryear{Landy \& Szalay}{1993}]{landy93}
Landy S., Szalay A., 1993, ApJ, 412, 64
\bibitem[\protect\citeauthoryear{Le Delliou et al.}{2006}]{delliou06}
Le Delliou M., Lacey C. G., Baugh C. M., Morris S. L., 2006, MNRAS, 365, 712
\bibitem[\protect\citeauthoryear{Li \& Draine}{2001}]{li01}
Li A., Draine B. T., 2001, ApJ, 554, 778
\bibitem[\protect\citeauthoryear{Li et al.}{2006}]{li06}
Li C., Kauffmann G., Jing Y. P., White S. D. M., B\"{o}rner G., Cheng F. Z.,
2006, MNRAS, 368, 21
\bibitem[\protect\citeauthoryear{Limber}{1953}]{limber53}
Limber D. N., 1953, ApJ, 117, 134
\bibitem[\protect\citeauthoryear{Maddox et al.}{2010}]{maddox10}
Maddox S. J., et al., 2010, A\&A, 518, 11
\bibitem[\protect\citeauthoryear{Michalowski et al.}{2010}]{michalowski09}
Michalowski M. J., Hjorth J., Watson D., 2010, A\&A, 514, 67
\bibitem[\protect\citeauthoryear{Mo \& White}{1996}]{mo96}
Mo H. J., White S. D. M, 1996, MNRAS, 282, 347
\bibitem[\protect\citeauthoryear{Meneux et al.}{2008}]{meneux08}
Meneux B., et al., 2008, A\&A, 478, 299
\bibitem[Nagashima et al.(2005a)]{nagashima05a} 
Nagashima, M., Lacey, 
C.~G., Baugh, C.~M., Frenk, C.~S., \& Cole, S.\ 2005, MNRAS, 358, 1247 
\bibitem[\protect\citeauthoryear{Nagashima et al.}{2005b}]{nagashima05b}
Nagashima M., Lacey C. G., Okamoto T., Baugh C. M., Frenk C. S., Cole S., 2005, MNRAS, 363, 31
\bibitem[\protect\citeauthoryear{Norberg et al.}{2001}]{norberg01}
Norberg P., et al., 2001, MNRAS, 328, 64
\bibitem[\protect\citeauthoryear{Orsi et al.}{2008}]{orsi08}
Orsi A., Lacey C. G., Baugh C. M., Infante L., 2008, MNRAS, 391, 1589
\bibitem[\protect\citeauthoryear{Panuzzo et al.}{2007}]{panuzzo07}
Panuzzo P., Granato G. L., Buat V., Inoue A. K., Silva L., Iglesias-P\'{a}ramo J., Bressan A.,
2007, MNRAS, 375, 640
\bibitem[\protect\citeauthoryear{Peebles}{1980}]{peebles80}
Peebles P. J. E., 1980, {\itshape The large-scale structure of the Universe}, Princeton, N.J.,
Princeton University Press
\bibitem[\protect\citeauthoryear{Percival et al.}{2003}]{percival03}
Percival W. J., Scott D., Peacock J. A., Dunlop J. S., 2003, MNRAS, 338, 31
\bibitem[\protect\citeauthoryear{Pope et al.}{2008}]{pope08}
Pope et al., 2008, ApJ, 675, 1171
\bibitem[\protect\citeauthoryear{Power et al.}{2010}]{power10}
Power C., Baugh C. M., Lacey C. G, 2010, MNRAS, 406, 43
\bibitem[\protect\citeauthoryear{Riedmiller \& Braun}{1993}]{rprop}
Riedmiller M., Braun H., 1993, Proc. of the IEEE Intl. Conf. on Neural Networks, 586
\bibitem[\protect\citeauthoryear{Schurer et al.}{2009}]{schurer09}
Schurer A., Calura F., Silva L., Pipino A., Granato G. L., Matteucci F., Maiolino R.,
2009, MNRAS, 394, 2001
\bibitem[\protect\citeauthoryear{Scott et al.}{2002}]{scott02}
Scott S. E., et al., 2002, MNRAS, 331, 817
\bibitem[Scott et al.(2006)]{scott06} Scott, S.~E., Dunlop, 
J.~S., \& Serjeant, S.\ 2006, MNRAS, 370, 1057 
\bibitem[\protect\citeauthoryear{Seljak}{2000}]{seljak00}
Seljak U., 2000, MNRAS, 318, 203
\bibitem[\protect\citeauthoryear{Sheth, Mo \& Tormen}{2001}]{smt01}
Sheth R. K., Mo H. J., Tormen G., 2001, MNRAS, 323, 1
\bibitem[\protect\citeauthoryear{Silva et al.}{1998}]{silva98}
Silva L., Granato G. L., Bressan A., Danese L., 1998, ApJ, 509, 103
\bibitem[\protect\citeauthoryear{Silva et al.}{2010}]{silva10}
Silva L., Schurer A., Granato G. L., Almeida C., Baugh C. M., Frenk C. S., 
Lacey C. G., Paoletti L., Petrella A., Selvestrel D., 2010, MNRAS, submitted, (arXiv:1006.4637)
\bibitem[\protect\citeauthoryear{Simon}{2007}]{simon07}
Simon P., 2007, A\&A, 473, 711
\bibitem[\protect\citeauthoryear{Smail et al.}{1997}]{smail97}
Smail I., Ivison R. J., Blain A. W., 1997, ApJ, 490, L5
\bibitem[\protect\citeauthoryear{Smail et al.}{2002}]{smail02}
Smail I., Ivison R. J., Blain A. W., Kneib J. -P., 2002, MNRAS, 331, 495
\bibitem[\protect\citeauthoryear{Solomon \& Vanden Bout}{2005}]{solomon05}
Solomon P. M., Vanden Bout P. A., 2005, ARA\&A, 43, 677
\bibitem[\protect\citeauthoryear{Springel et al.}{2005}]{springel}
Springel V., et al., 2005, Nature, 435, 629
\bibitem[\protect\citeauthoryear{Swinbank et al.}{2008}]{swinbank}
Swinbank A. M., et al., 2008, MNRAS, 391, 420
\bibitem[\protect\citeauthoryear{Tacconi et al.}{2006}]{tacconi06}
Tacconi L. J., et al., 2006, ApJ, 640, 228
\bibitem[\protect\citeauthoryear{van Kampen et al.}{2005}]{vanKampen05}
van Kampen E., et al., 2005, MNRAS, 359, 469
\bibitem[\protect\citeauthoryear{Vega et al.}{2005}]{vega05}
Vega O., Silva L., Panuzzo P., Bressan A., Granato G. L., Chavez M.,
2005, MNRAS, 364, 1286
\bibitem[\protect\citeauthoryear{Wei\ss{} et al.}{2009}]{weiss09}
Wei\ss{} A., et al., 2009, ApJ, 707, 1201
\bibitem[\protect\citeauthoryear{Wilman et al.}{2005}]{wilman05}
Wilman R. J., Gerssen J., Bower R. G., Morris S. L., Bacon R., de Zeeuw P. T., Davies R. L.,
2005, Nature, 436, 227
\bibitem[\protect\citeauthoryear{Zehavi et al.}{2005}]{zehavi05}
Zehavi I., et~al., 2005, ApJ, 621, 22
}

\end{thebibliography}
\end{document}